\documentclass{article}

\usepackage{arxiv}

\usepackage[utf8]{inputenc} 
\usepackage[T1]{fontenc}    
\usepackage[numbers,sort&compress]{natbib}
\usepackage{hyperref}       
\usepackage{url}            
\usepackage{booktabs}       
\usepackage{amsfonts}       
\usepackage{nicefrac}       
\usepackage{color,xcolor}
\usepackage{microtype}      
\usepackage{graphicx}
\usepackage{amssymb}
\usepackage{amsmath}
\usepackage{subcaption}
\usepackage{booktabs}

\DeclareMathOperator\erf{erf}
\newcommand{\vect}[1]{\boldsymbol{{#1}}}

\title{Parameter identifiability in PDE models of fluorescence recovery after photobleaching}

\author{
   Maria-Veronica Ciocanel \\
  Department of Mathematics and Department of Biology \\
  Duke University \\
   \And
  Lee Ding\\
  Department of Biostatistics\\
  Harvard University\\
  \And
   Lucas Mastromatteo \\
   GlaxoSmithKline\\
  \And
  Sarah Reichheld \\
  Department of Neuroscience\\
  Brown University
  \And
  Sarah Cabral \\
  Remix Therapeutics\\
  \And
  Kimberly Mowry \\
  Department of Molecular Biology, Cell Biology \& Biochemistry\\
  Brown University
\And
  Bj\"{o}rn Sandstede \\
  Division of Applied Mathematics \\
  Brown University
}

\begin{document}
\maketitle

\begin{abstract}
Identifying unique parameters for mathematical models describing biological data can be challenging and often impossible. Parameter identifiability for partial differential equations models in cell biology is especially difficult given that many established \textit{in vivo} measurements of protein dynamics average out the spatial dimensions. Here, we are motivated by recent experiments on the binding dynamics of the RNA-binding protein PTBP3 in RNP granules of frog oocytes based on fluorescence recovery after photobleaching (FRAP) measurements. FRAP is a widely-used experimental technique for probing protein dynamics in living cells, and is often modeled using simple reaction-diffusion models of the protein dynamics. We show that current methods of structural and practical parameter identifiability provide limited insights into identifiability of kinetic parameters for these PDE models and spatially-averaged FRAP data. We thus propose a pipeline for assessing parameter identifiability and for learning parameter combinations based on re-parametrization and profile likelihoods analysis. We show that this method is able to recover parameter combinations for synthetic FRAP datasets and investigate its application to real experimental data.
\end{abstract}

\keywords{parameter identifiability, partial differential equations, profile likelihood, FRAP, RNA binding proteins}

\section{Introduction}\label{sec:intro}

Many mathematical models of biological processes aim to test relevant biological mechanisms, which are characterized using parameters. Estimating the underlying parameters helps connect and validate mathematical models with existing measurements and thus provide insights into mechanistic understanding of the biological process. However, mathematical models can suffer from identifiability issues, meaning that it may not be possible to uniquely determine the model parameters from the available data. Identifiability is thus a crucial problem in parameter estimation, and various approaches from statistics, applied mathematics, and engineering have been devised to address it \cite{ollivier1990probleme,ljung1994global,evans2000extensions,audoly2001global,hengl2007data,gutenkunst2007universally,chis2011structural,hong2020global}. Identifiability problems are typically categorized into structural identifiability, which involves issues arising from the model structure alone, and practical identifiability, which involves issues with parameter estimation stemming from the incorporation of real and noisy data \cite{eisenberg2014determining}.

In mathematical biology, many of these approaches have been more extensively tested and used in models of epidemic and disease treatment dynamics or in systems biology models \cite{renardy2022structural}. For example, \cite{cobelli1980parameter,gutenkunst2007universally,miao2011identifiability,chis2011structural} review theoretical results and algorithms for structural and practical identifiability of linear and nonlinear ordinary differential equations (ODE) models, with applications to disease dynamics and systems biology processes. 

In the study of macromolecular dynamics inside cells, spatial movement - characterized by diffusion, transport, and binding dynamics - can be significant and has an impact on the parameters that describe a given model. As a result, partial differential equations (PDEs) that incorporate the dynamics of proteins as a function of time and space are often an appropriate modeling framework. However, PDEs present challenges when studying identifiability measures, since these equations have more variables, contain derivatives, and include boundary conditions \cite{renardy2022structural}. Fewer studies have thus dealt with identifiability for PDE models \cite{renardy2022structural,browning2023structural}. 

\textblue{Here, we focus on biological data obtained from a versatile experimental technique for probing protein dynamics in living cells: FRAP (fluorescence recovery after photobleaching). Our prior results suggest} that parameters in PDE systems that model data obtained from  \textblue{FRAP experiments} may not be identifiable. \textblue{In particular, when estimating kinetic parameters describing mRNA dynamics based on FRAP data, we observed that the predicted parameters (especially binding rates) could vary across orders of magnitude for the same experimental settings \cite{ciocanel2017analysis}. Similarly, the theoretical studies in \cite{sprague2004analysis,alexander2022inferences} have found that only specific model parameters could be identified
from FRAP data in certain parameter regimes. These identifiability issues with estimating kinetic parameters from FRAP data can thus result} in potentially spurious predictions of cellular dynamics.

\textblue{We are particularly} motivated by questions surrounding protein and RNA dynamics in \textit{Xenopus laevis} frog oocytes \cite{neil2021bodies,cabral2022multivalent}. Proteins and RNAs organize into membraneless compartments called \textblue{biomolecular condensates (also called RNP granules,} localization bodies, or L-bodies) in the developing oocytes. \textblue{FRAP has been playing a key role as a technique to study newly-discovered biological processes such as the formation and organization of biomolecular condensates \cite{kenworthy2023s}.} We consider a reaction-diffusion PDE model of FRAP microscopy experiments for the dynamics of an RNA-binding protein that is enriched in \textblue{RNP} granules and investigate the limitations of existing structural and practical identifiability techniques for this model and data. We propose \textblue{an alternative} custom pipeline for extracting identifiable parameter combinations for the PDE model based on time-series FRAP data. \textblue{This approach allows the prediction of the protein diffusion coefficient and of the relationship between binding and unbinding rates of the RNA-binding protein.} We illustrate the application of this framework for both synthetic and experimental FRAP datasets. Given additional biological information on relevant binding rate parameters, this approach may allow the inference of all individual model parameters.

\section{Biological motivation and fluorescence microscopy data}\label{sec:bio_motivation}

RNP granules are membraneless compartments containing RNA and other proteins, serving diverse biological functions. In developing \textit{Xenopus laevis} oocytes, maternal mRNAs are packaged into large RNP granules that localize to specific subcellular locations, in a process that is required for embryonic patterning \cite{neil2021bodies} \textblue{(see Figure~\ref{fig:postbleach_fit}A)}. The assembly of RNAs into these RNP granules (termed localization bodies or L-bodies) requires the interaction of RNAs with RNA-binding proteins (RBPs). The data suggest that the protein dynamics are influenced by the strength and number of interactions of RBPs with the non-dynamic RNA in L-bodies \cite{cabral2022multivalent}. An example of a multivalent RNA-binding protein is PTBP3, which is highly co-localized with L-bodies in \textit{Xenopus laevis} oocytes \cite{cabral2022multivalent}. PTBP3 has four domains (termed RRM1, RRM2, RRM3, and RRM4) that can bind to RNA, making it an ideal model for studying the strength of interactions within L-bodies. In particular, experimental manipulations in this system can generate PTBP3 RNA-binding mutants, where the ability of one or more RNA-binding domains to bind to RNA is abolished \cite{cabral2022multivalent}. Quantifying the binding of PTBP3, and its mutants, to RNA would therefore be useful in contributing to our understanding of how protein dynamics are regulated in L-bodies and other RNP granules.
An important experimental technique for assessing protein dynamics \textit{in vivo} is fluorescence recovery after photobleaching (FRAP). FRAP is a well-established approach to studying the binding and diffusion of molecules in cells \cite{kenworthy2023s}. \textblue{FRAP is also considered to be one of the most versatile methods of studying protein dynamics and binding characteristics in living cells \cite{ishikawa2012advanced}.} This microscopy experiment relies on bleaching a small region in a cell expressing a fluorescent protein or nucleic acid, and quantifying the recovery of fluorescence in that bleach spot over time \textblue{(see Figure~\ref{fig:postbleach_fit}A,B)}. \textblue{The output of the FRAP experiment consists of a series of images (such as Figure~\ref{fig:postbleach_fit}C) for each time point. These images are then used to calculate the amount of fluorescence intensity inside the bleached region as a function of time. It is this time series of real-valued fluorescence intensities that is then used in subsequent analysis, and we refer to Figure~\ref{fig:PL_s_mut3_expt}C for a sample FRAP intensity curve. We emphasize that the spatial pixel by pixel information captured in FRAP experiments is typically not considered to be robust enough for analysis. The analysis of FRAP data therefore focuses on the time series that tracks the overall integrated fluorescence recovery in the bleached region to gain insight into the dynamic processes that the proteins undergo \cite{ishikawa2012advanced}.}

In this work, we use the FRAP experimental measurements in \cite{cabral2022multivalent} in order to determine parameter regimes of interest. These FRAP datasets consist of fluorescence recovery curves that are adjusted to correct for photobleaching during image acquisition, as we previously outlined in \cite{powrie2016using}. In these experiments, the fluorescence in the bleach spot (a square with side $l=10\mu$m) is recorded at 5-second intervals for a total of $500$ seconds. 

\section{Mathematical modeling of FRAP}\label{sec:math_model}

\subsection{PDE model of protein dynamics}\label{sec:pde_model}

We model the dynamics of PTBP3 using a system of linear reaction-diffusion PDEs. The variables we study correspond to concentrations of PTBP3 in different dynamical states: $f(x,y,t)$ denotes the concentration of free protein and $c(x,y,t)$ refers to the concentration of bound complexes at location $(x,y)$ and time $t$. We assume PTBP3 can transition between the diffusing and stationary states, so that the dynamics is described by the PDE system:
\begin{align}\label{eq:diff_pause}
\frac{\partial f}{\partial t} & = D \Delta f -\beta_2 f + \beta_1 c \,, \nonumber\\
\frac{\partial c}{\partial t} & = \beta_2 f - \beta_1 c\,,
\end{align}
where $D$ denotes the diffusion constant in the diffusing state, $\beta_1$ is the rate of transition from the stationary to the diffusing state, and $\beta_2$ is the rate of transition from the diffusing to the stationary state. This model is equivalent to the reaction-diffusion system we previously studied in \cite{ciocanel2017analysis} for non-localizing RNA dynamics and has also been previously used and analyzed in other works \textblue{on quantifying FRAP experiments}, including \cite{sprague2004analysis}. Our goal is to estimate the reaction rate parameters $\beta_1$ and $\beta_2$ and the diffusion constant $D$ from experimental FRAP data. \textblue{It is worth noting that RNA and RNA-binding protein dynamics are not modelled explicitly in these equations. The binding of PTBP3 into complexes (i.e., transition rate $\beta_2$) likely depends on the spatial organization of RNAs and other binding proteins in the L-bodies, which remains challenging to investigate.}

A key assumption underlying this model is that the binding interactions of PTBP3 involve a single binding state. Four binding domains have been identified for PTBP3, of which two were shown to bind to the non-dynamic L-body RNA \cite{cabral2022multivalent}. Mathematical models involving multiple independent binding sites are more challenging to evaluate due to the increased dimension of the parameter space, and generally show similar FRAP behaviors \cite{sprague2004analysis}. We therefore proceed with the simplifying assumption of a single binding site for the PTBP3 reaction. We comment on the limitations of this assumption in the Discussion.

\subsection{Postbleach intensity profile model}\label{sec:ic_postbleach}

\begin{figure}[t]
\center
\includegraphics[width=\textwidth]{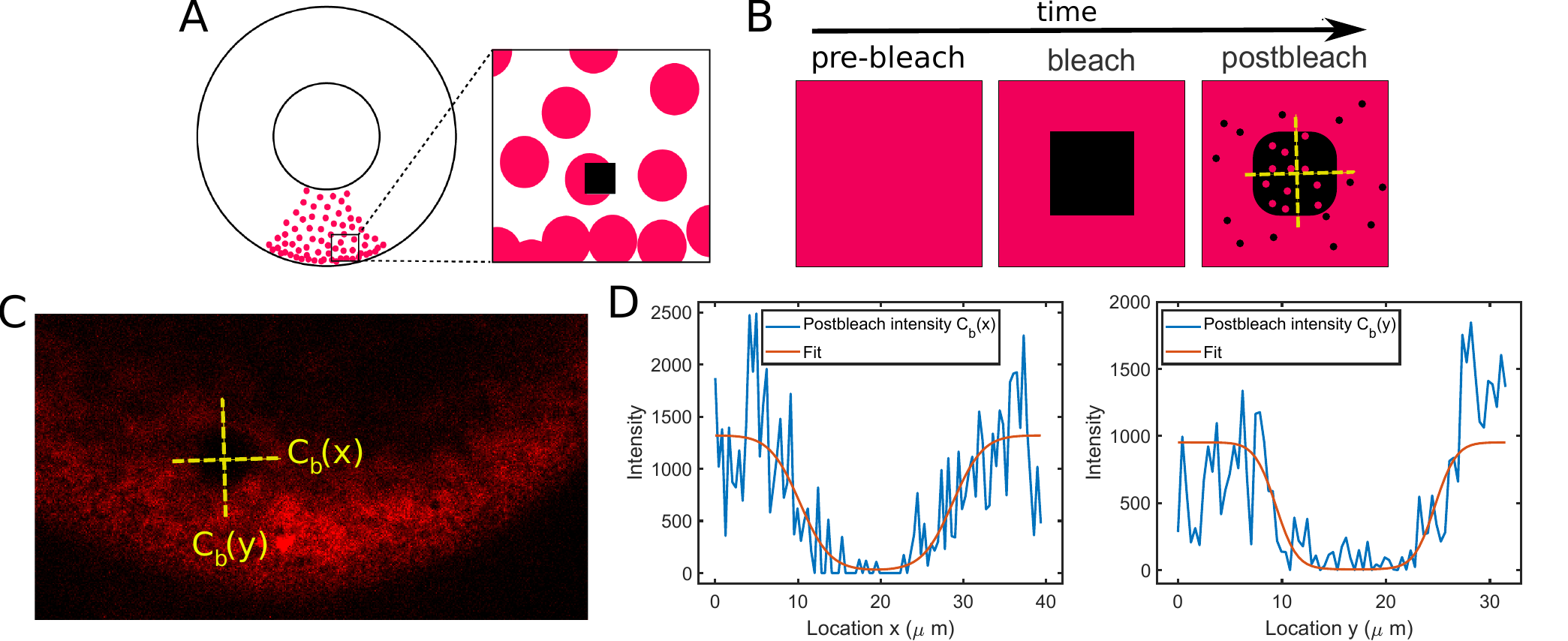}
\caption{\textblue{A) Schematic of a stage II Xenopus oocyte with RNA granules localizing at the vegetal cortex (bottom) shown in magenta. The black square region is shown magnified on the right, with a cartoon of a FRAP bleach spot. B) The timeline of Fluorescence Recovery After Photobleaching (FRAP) shows bleaching of a small square region (at the bleach time) in a previously-fluoresced region of the cell (at the pre-bleach time). The first postbleach time already shows that non-bleached and fluorescent molecules mix between the fluoresced and bleached regions.} C) An image of the vegetal cytoplasm of a \textit{Xenopus laevis} oocyte expressing fluorescently-labeled PTBP3 (red) in L-bodies is shown, with a $10~\mu$m photobleach square ROI. \textblue{This image corresponds to the postbleach time point in the cartoon in B}. Yellow dashed lines show sample extraction of the fluorescence profiles $C_b(x)$ and $C_b(y)$ in the $x$ and $y$ directions from the postbleach intensity data, \textblue{as shown in panel B as well}. D) Fitted fluorescence postbleach profiles along the $x$ and $y$ directions. \textblue{The estimated parameters are $\alpha_x=2.33$, $r_x=5.64$ (with $R^2=0.39$) and $\alpha_y=2.72$, $r_y=3.27$ (with $R^2=0.37$).}}
\label{fig:postbleach_fit}
\end{figure}

To determine initial conditions for the concentrations of PTBP3 in the PDE model~\eqref{eq:diff_pause}, we consider a model of the experimental FRAP postbleach intensity profiles on the focal plane of the fluorescence distribution \cite{ciocanel2017analysis}. The photobleaching process in FRAP is usually assumed to be an irreversible first-order reaction of the form
\begin{align}
\frac{\partial C_b}{\partial t}(x,y,t) &= -\alpha K(x,y) C_b(x,y,t)
\end{align}
for the fluorophore concentration $C_b(x,y,t)$, where $\alpha$ is a bleaching parameter and $K(x,y)$ is the effective bleaching intensity distribution. Since the initial condition of model~\eqref{eq:diff_pause} corresponds to the spatial concentration of fluorophores at the first postbleach time \textblue{(see Figure~\ref{fig:postbleach_fit}B)}, we therefore seek:
\begin{align} \label{eq:ic}
C_b(x,y,0) &= C_0 \mathrm{e}^{-\alpha K(x,y)}\,.
\end{align}

The FRAP experiments in \cite{cabral2022multivalent} use square bleach regions of interest (ROIs). We therefore adapt the approach in \cite{deschout2010straightforward}, which considers a rectangular FRAP bleach spot. The effective bleaching intensity distribution $K(x,y)$ is calculated as the convolution of the bleach geometry $B(x,y)$ and the time-averaged bleaching intensity distribution $\langle I_b (x-x',y-y',t) \rangle$:
\begin{align}
K(x,y) = \int_{-\infty}^{\infty} \int_{-\infty}^{\infty}  B(x',y') \langle I_b (x-x',y-y',t) \rangle \,\mathrm{d}x'\,\mathrm{d}y'\,.
\end{align}
We assume a square photobleach area with side length $l$ and a Gaussian photobleaching intensity distribution \cite{deschout2010straightforward}:
\begin{align}
B(x,y) &= \begin{cases}
      1, & \text{if}\ \lvert x\rvert<l/2 \text{ and } \lvert y \rvert <l/2\\
      0, & \text{otherwise}
    \end{cases} \,, \label{eq:Bxy} \\
\langle I_b (x,y,t) \rangle &= I_0 \mathrm{e}^{-2\frac{x^2+y^2}{r^2}}\,,
\end{align}
where $r$ is the effective radius of the distribution. 

We therefore obtain for the effective bleaching intensity distribution:
\begin{align}
K(x,y) &= I_0 \int_{-l/2}^{l/2} \mathrm{e}^{-\frac{(x-x')^2}{r^2}} \,\mathrm{d}x' \int_{-l/2}^{l/2} \mathrm{e}^{-\frac{(y-y')^2}{r^2}}\,\mathrm{d}y' \,,\\
           &= I_0 \int_{(x-l/2)/r}^{(x+l/2)/r} \mathrm{e}^{-u^2} \,\mathrm{d}u \int_{(y-l/2)/r}^{(y+l/2)/r} \mathrm{e}^{-v^2}\,\mathrm{d}v \,,\\
           &= \tilde{I}_0 \left[ \erf\left(\frac{x+l/2}{r}\right) -  \erf\left(\frac{x-l/2}{r}\right)  \right] \left[ \erf\left(\frac{y+l/2}{r}\right) -  \erf\left(\frac{y-l/2}{r}\right)  \right]\,.
\end{align}
Plugging this into \eqref{eq:ic} for the initial fluorophore concentration yields: 
\begin{align}\label{eq:Cb_full}
C_b(x,y) &= C_0 \mathrm{e}^{-\tilde{\alpha} \left[ \erf\left(\frac{x+l/2}{r}\right) -  \erf\left(\frac{x-l/2}{r}\right)  \right]  \left[ \erf\left(\frac{y+l/2}{r}\right) -  \erf\left(\frac{y-l/2}{r}\right)  \right] }\,.
\end{align} 

Since the experimental postbleach profiles show some asymmetry along the two spatial dimensions (see Figure~\ref{fig:postbleach_fit}C), we extract fluorescence profiles $C_b(x)$ and $C_b(y)$ in the $x$ and $y$ directions from the postbleach intensity data and fit them to expressions of the form:
\begin{align}
C_b(x) &= C_x \mathrm{e}^{-\alpha_x \left[ \erf\left(\frac{x+l/2}{r_x}\right) -  \erf\left(\frac{x-l/2}{r_x}\right)  \right] } \,, \label{eq:Cb_x}\\
C_b(y) &= C_y \mathrm{e}^{-\alpha_y \left[ \erf\left(\frac{y+l/2}{r_y}\right) -  \erf\left(\frac{y-l/2}{r_y}\right)  \right] }\,. \label{eq:Cb_y}
\end{align} 
In particular, we estimate the parameters $r_x$ and $\alpha_x$ by fitting the fluorescence profile $C_b(x)$ to equation~\eqref{eq:Cb_x} and parameters $r_y$ and $\alpha_y$ by fitting the fluorescence profile $C_b(y)$ to equation~\eqref{eq:Cb_y} using standard nonlinear least-squares estimation in Matlab using the function \texttt{nlinfit}. Here we use $l=10~\mu$m, consistent with the experiments in \cite{cabral2022multivalent}. We illustrate sample postbleach intensity profiles and the corresponding fitted curves in Figure~\ref{fig:postbleach_fit}D. \textblue{The coefficient of determination for the fit to the fluorescence profile $C_b(x)$ is $R^2=0.39$, and for the fit to the fluorescence profile $C_b(y)$ is $R^2=0.37$. We note that the derived models~\eqref{eq:Cb_x} and~\eqref{eq:Cb_y} for the postbleach intensity fit the protein distribution data in Figure~\ref{fig:postbleach_fit}C well, particularly in the locations corresponding to the photobleached region. This model cannot however account for the variation due to the noisy fluorescence in the rest of the oocyte (as illustrated by the edges of the spatial domain in Figure~\ref{fig:postbleach_fit}C,D).} 

Since the estimated $\alpha_x$ and $\alpha_y$ parameter values are very similar for all datasets considered, we use the following final form for the initial fluorophore concentration:
\begin{align}\label{eq:Cb_full_approx}
C_b(x,y) &\sim C_0 \mathrm{e}^{-\tilde{\alpha} \left[ \erf\left(\frac{x+l/2}{r_x}\right) -  \erf\left(\frac{x-l/2}{r_x}\right)  \right]  \left[ \erf\left(\frac{y+l/2}{r_y}\right) -  \erf\left(\frac{y-l/2}{r_y}\right)  \right] }\,.
\end{align} 
\textblue{We note that parameter $C_0$ acts as a normalization constant for the level of the intensity profile. This parameter is fitted here, but it is also estimated in the final optimization of the parameters of interest, which describe the protein dynamics (see Section~\ref{sec:par_estimation}).}

Finally, the initial conditions for the model equations \eqref{eq:diff_pause} are given by:
\begin{align}\label{eq:IC_final}
f(x,y,t=0) & = p C_b(x,y) \,, \nonumber\\
c(x,y,t=0) & = (1-p) C_b(x,y) \,,
\end{align}
where the initial postbleach profile $C_b(x,y)$ is given in \eqref{eq:Cb_full_approx} and the parameter $p \in [0,1]$ denotes the initial fraction of PTBP3 protein in the diffusing state, which we will also determine from the data as described below. As shown in \cite{ciocanel2017analysis}, parameter estimation for FRAP experiments is sensitive to the initial condition given by the postbleach profile. We therefore use these data-informed initial conditions for all the studies carried out in this work.

\subsection{Deterministic parameter estimation}\label{sec:par_estimation}

In testing the techniques proposed here, we consider both synthetic and experimental FRAP data. The experimental fluorescence intensity data is collected in \cite{cabral2022multivalent} at every $5$s intervals up to a total time of $500$s. We adjust the microscopy data by correcting for background fluorescence and dividing the resulting fluorescence recovery by the intensity of a neighboring ROI for each time point, as we previously described in \cite{powrie2016using}. 

We denote the real FRAP data by $\mathrm{FRAP}_\mathrm{true}(t)$. The corresponding quantity from the FRAP model described in Section~\ref{sec:pde_model} is then denoted by $\mathrm{FRAP}(t,\vect{\theta})$ and calculated as
\begin{align}\label{eq:FRAP_curve}
\mathrm{FRAP}(t,\vect{\theta}) & = \int_{-l/2}^{l/2}\int_{-l/2}^{l/2} (f+c)(x,y,t,\vect{\theta}) \,\mathrm{d}x \,\mathrm{d}y \,.
\end{align}
Here, $\vect{\theta}$ is the vector of parameters of interest and $l$ is the side of the square bleach ROI. We let $\vect{\theta}=(D,\beta_1,\beta_2,p,C_0)$ and note that $D,\beta_1,\beta_2$ are kinetic parameters describing the dynamics of PTBP3 proteins (equations~\eqref{eq:diff_pause}), while $p$ and $C_0$ are parameters that describe the initial postbleach profile in each protein population (equations~\eqref{eq:Cb_full_approx} and \eqref{eq:IC_final}). \textblue{We note that $p$ is the initial fraction of proteins in the diffusing state and is thus unitless. $C_0$ has units of concentration (see equation~\eqref{eq:ic}) and acts as a normalization constant for the initial fluorophore concentration. Parameters $p$ and $C_0$ cannot be validated with data and do not yield significant insights into the protein dynamics; we are therefore most interested in the estimation and identifiability of kinetic parameters $\{D,\beta_1,\beta_2\}$.}

As in \cite{ciocanel2017analysis}, we numerically integrate equations~\eqref{eq:diff_pause} using an efficient exponential time-differencing fourth-order Runge-Kutta scheme \cite{cox2002exponential,kassam2005fourth} for time integration coupled with Fourier spectral methods for space discretization to solve for $\mathrm{FRAP}(t,\vect{\theta})$. \textblue{The settings for the numerical solution of the PDE system are the same as outlined in \cite[Supplementary Material S5]{ciocanel2017analysis}, with the exception that here we use a larger $100~\mu$m $\times$ $100~\mu$m spatial domain. Code for implementing this numerical scheme is available in the Github repository \cite{GithubFRAP} associated with \cite{ciocanel2017analysis}.} We then use the MATLAB optimization routine \texttt{lsqnonlin} to determine the parameter set that minimizes the $L^2$-norm difference between the true and model FRAP curves \textblue{on the time interval $I$}:
\begin{align}\label{eq:min_theta}
\vect{\hat{\theta}} & = \textblue{\arg}\min_{\vect{\theta}} \| \mathrm{FRAP}_\mathrm{true}(\cdot)- \mathrm{FRAP}(\cdot,\vect{\theta}) \|_{\textblue{L^2(I,\mathbb{R})}}^2 \,.
\end{align}

We previously found that the initial guesses for parameters describing FRAP dynamics can be key in ensuring convergence in deterministic parameter estimation for this type of data \cite{ciocanel2017analysis}. \textblue{In addition, since the data are obtained through \textit{in vivo} cell measurements, there is little prior knowledge on the magnitudes of the kinetic parameters of interest. We therefore use a similar approach as employed in prior work \cite{sprague2004analysis,ciocanel2017analysis} and} we carry out parameter sweeps that sample through \textblue{a range of} values of $D,\beta_1,\beta_2,p$. \textblue{Each such parameter combination includes a value of $D\in\{10^{-3},5\times 10^{-3}, 10^{-2}, 5\times 10^{-2}, 10^{-1}, 5\times 10^{-1}, 1,1.5,2\}$, a value of $\beta_1 = 10^{n_1}$ with $n_1\in\{-6,-5,...,1,2\}$, a value of $\beta_2 = 10^{n_2}$ with $n_2\in\{-6,-5,...,1,2\}$, and a value of $p \in \{0, 0.25, 0.5, 0.75,1\}$. For each parameter combination in this sweep, we generate FRAP curves using the PDE model \eqref{eq:diff_pause} for the given parameters and evaluate} the $L^2$-norm difference between the generated FRAP curves and the true data. \textblue{The sweeps thus require many forward evaluations of PDE model~\eqref{eq:diff_pause} with the efficient numerical scheme outlined above. We summarize this step in the flowchart Figure~\ref{fig:flowchart}A.} 

We then choose the parameter sets that yield the smallest differences \textblue{between data and generated curves to serve as} initial guesses \textblue{for the rough magnitudes of the parameters. These initial guesses are then refined using} the optimization routine \textblue{\textblue{lsqnonlin}} to estimate specific values of $\vect{\theta}=(D,\beta_1,\beta_2,p,C_0)$. \textblue{We keep the normalization parameter $C_0$ fixed throughout the parameter sweeps since the initial guess for this parameter is informed by fitting the initial point on the FRAP curve to the form of the initial fluorophore concentration in \eqref{eq:Cb_full_approx}. This also allows to reduce the computational cost of these forward runs of the model. In the full optimization of parameters of the model, we use initial guesses for parameters informed from the parameter sweeps, and we allow $C_0$ to be estimated as well. This optimization of the model parameters is summarized in the flowchart Figure~\ref{fig:flowchart}B.}

We apply this framework to FRAP recovery data from experiments in stage II oocytes that test the dynamics and interactions of PTBP3 with specific RNA Recognition Motifs (RRMs) in L-bodies \cite{cabral2022multivalent}. In particular, we apply deterministic parameter estimation to experiments with wild-type PTBP3 (WT, Set~1), single RRM mutant PTBP3 (mut3 in \cite{cabral2022multivalent}, Set~2), and double RRM mutant PTBP3 (mut34 in \cite{cabral2022multivalent}, Set~3). The estimated kinetic parameters for the dynamics of PTBP3 are given in Table~\ref{tab:par_estimates} for several FRAP datasets from \cite{cabral2022multivalent}. \textblue{The resulting squared norm of the residual from each fit is given in the last column of Table~\ref{tab:par_estimates}. This quantity corresponds to $\| \mathrm{FRAP}_\mathrm{true}(\cdot)- \mathrm{FRAP}(\cdot,\vect{\theta}) \|_{\textblue{L^2(I,\mathbb{R})}}^2$, or the \texttt{resnorm} output from the Matlab optimization routine \texttt{lsqnonlin}. }

\begin{table}[ht]
\begin{center}
\caption{Results of deterministic parameter estimation for several wild-type (WT) and mutant PTBP3 FRAP datasets from \cite{cabral2022multivalent}.}\label{tab:par_estimates}%
\begin{tabular}{@{}llllllll@{}}
\toprule
Cell & PTBP3 & $D$  & $\beta_1$ & $\beta_2$ & $p$ & $C_0$ & \textblue{squared norm}\\
 & Type & [$\mu$m$^2$/s]  & [1/s] & [1/s] & \textblue{(unitless)} & \textblue{(conc.)} & \textblue{of residual}\\
\midrule
1    & WT & $0.26$   & $7.6\times10^{-5}$  & $9.6\times10^{-9}$  & 0.75 & 2.1 & 0.045\\
2    & WT & $0.22$   & $3.5\times10^{-3}$  & $2.2 \times 10^{-2}$  & 0.66 & 1.15 & 0.016\\
3    & WT & $0.84$   & $1.0\times10^{-3}$  & $3.1\times10^{-5}$  & 0.92 & 0.64 & 0.006 \\

1    & mut3 & $0.54$   & $9.4\times10^{-5}$  & $1.1\times10^{-9}$  & 0.83 & 2.7 & 0.022 \\
2    & mut3 & $0.56$   & $9.9\times 10^1$  & $7.2 \times10^{-1}$  & 0.23 & 0.5 & 0.045\\
3    & mut3 & $1.6$   & $5.5\times10^{-3}$  & $4.6\times10^{-4}$  & 0.31 & 0.79 & 0.044 \\

1    & mut34 & $1.93$   & $3.1\times10^{-5}$  & $5.8\times10^{-10}$  & 0.73 & 2.5 & 0.041 \\
2    & mut34 & $1.41$   & $2.5\times 10^{-5}$  & $4.7\times 10^{-8}$  & 0.82 & 3.2 & 0.029\\
3    & mut34 & $6.2$   & $9.6\times10^{-2}$  & $1.0\times 10^{-1}$  & 0.24 & 0.85 & 0.052 \\
\end{tabular}
\end{center}
\end{table}

\textblue{Even though the residuals for these fits are low, it} remains challenging to determine the level of confidence we can place in the results of the deterministic parameter estimation. Given the space-averaged FRAP data, it is very likely that some parameters of the spatio-temporal PDE model are not identifiable. As we have previously observed in \cite{ciocanel2017analysis}, while some parameters may show consistency within each wild-type or mutant setting, there is still wide variability, especially in the ranges of the reaction rates $\beta_1$ and $\beta_2$. In the following, we thus focus on synthetic data generated using PDE model~\eqref{eq:diff_pause} in order to investigate parameter identifiability of \textblue{the kinetic parameters} given FRAP data. 

In generating synthetic FRAP data, we consider three parameter regimes roughly inspired from the results of the parameter estimation procedure for the three wild-type and mutant settings (Sets~1-3). In addition, we consider an effective diffusion parameter regime (Set~4) as previously studied in \cite{sprague2004analysis} for equations~\eqref{eq:diff_pause}. In this regime, the reaction dynamics are much faster than diffusion, leading to rapid local equilibrium of the reaction process. This leads to FRAP recovery curves which can be characterized by the single parameter combination $D_{\mathrm{eff}} = \frac{D}{1+\beta_2/\beta_1}$ termed the \textit{effective diffusion coefficient} in \cite{sprague2004analysis}. We provide these parameter regimes in Table~\ref{tab:par_regimes}. These parameter values are used to generate \textblue{noiseless} synthetic FRAP data using model~\eqref{eq:diff_pause}, which we further use to assess parameter identifiability using established techniques in Section~\ref{sec:app_identifiability} and for benchmarking our proposed method in Section~\ref{sec:par_combinations}.

\begin{table}[ht]
\begin{center}
\caption{Set of parameter regimes chosen to generate synthetic FRAP data using model~\ref{eq:diff_pause}. The last two columns display the determinant and eigenvalues of the Fisher information matrix for each parameter set (Section~\ref{sec:structural} and \ref{sec:app_structural}).}\label{tab:par_regimes}%
\begin{tabular}{@{}llllllll@{}}
\toprule
 & $D$ [$\mu$m$^2$/s]  & $\beta_1$ [1/s] & $\beta_2$ [1/s] & $p$ & $C_0$ & $\det(F)$ & eig$(F)$ \\
\midrule
Set 1 & $0.1$  & $10^{-3}$  & $10^{-3}$  & 0.5  & 1.5  & $1.3\times 10^{-4}$  & $5\times 10^{-4},0.1,2.9$ \\
Set 2 & $1$    & $10^{-3}$  & $10^{-4}$  & 0.25 & 0.75 & $1.0\times 10^{-2}$  & $9\times 10^{-3},0.5,2.5$ \\
Set 3 & $2$    & $10^{-5}$  & $10^{-3}$  & 0.75 & 2    & $3.8\times 10^{-2}$  & $3\times 10^{-2},0.6,2.4$ \\
Set 4 & $0.1$  & $5$        & $5$        & 0.5  & 1.5  & $7.7\times 10^{-17}$ & $0, 0, 3$ \\
\end{tabular}
\end{center}
\end{table}

\section{Methods for practical and structural parameter identifiability}\label{sec:identifiability}

Our goal is to investigate the identifiability of kinetic parameters of FRAP models. We begin by reviewing established methods for practical and structural parameter identifiability in differential equations models. Throughout this work, the parameter learning and identifiability methods considered will apply more generally to a model of the form
\begin{align}\label{eq:general_model}
    \frac{\partial\vect{x}}{\partial t} &= f(\vect{x},t;\vect{\theta})\,,
\end{align}
where $\vect{\theta}=(\theta_1, \theta_2,...,\theta_n )$ is the vector of model parameters of interest and the model output is given by
\begin{align}\label{eq:general_model_output}
    y &= g(\vect{x},t;\vect{\theta})\,.
\end{align}
For our application, model~\eqref{eq:general_model} will consist of the partial differential equations in \eqref{eq:diff_pause}, and the model output $y$ will consist of the time-series FRAP measurements defined in \eqref{eq:FRAP_curve}. 

In this section, we provide a brief overview of established methods of parameter identifiability for models and output as in equations~\eqref{eq:general_model} and \eqref{eq:general_model_output}. As commonly done in studies of identifiability analysis, we distinguish between structural identifiability, which considers issues with identifying parameters based on the model structure alone, and practical identifiability, which considers issues that stem from identification based on real and potentially noisy data \cite{eisenberg2014determining}.

\subsection{Structural identifiability}\label{sec:structural}

An established technique for assessing the local structural identifiability of a model consists of constructing the Fisher Information Matrix (FIM). This matrix, denoted by $F$, captures the amount of information contained in the model output $y(\vect{t})$ about the set of parameters $\vect{\theta}$ \cite{eisenberg2014determining}. Here we assume that the data measurements are available at times $\vect{t} = (t_1, t_2, ..., t_m)$. Based on the concept of sensitivity identifiability introduced in \cite{reid1977structural} and reviewed in \cite{cobelli1980parameter}, this technique requires calculating the sensitivity matrix:
\begin{align*}
    X &= \begin{pmatrix} \frac{\partial{y}}{\partial{\theta_1}}(\vect{t};\vect{\theta^0}) & \frac{\partial{y}}{\partial{\theta_2}}(\vect{t};\vect{\theta^0}) & \ldots & \frac{\partial{y}}{\partial{\theta_n}}(\vect{t};\vect{\theta^0})  \end{pmatrix}\,,
\end{align*}
where $\vect{\theta^0}$ is a set of baseline parameters around which the sensitivities are evaluated. The Fisher Information Matrix is then given by the symmetric $n \times n$ matrix $F = X^T X$.
Studies \cite{reid1977structural,cobelli1980parameter} show that identifiability of the parameter set $\vect{\theta}$ requires nonsingularity of matrix $F$. In practice, the parameter sensitivities $\frac{\partial{y}}{\partial{\theta_i}}(\vect{t};\vect{\theta^0})$ are approximated numerically, and the parameter set $\vect{\theta}$ is considered unidentifiable when $\det(F)$ is small \cite{eisenberg2014determining}. The rank of the matrix $F$ \textblue{(or equivalently, the number of non-zero eigenvalues of symmetric matrix $F$)} gives the number of identifiable parameter combinations \cite{cobelli1980parameter,cintron2009sensitivity,eisenberg2014determining}, but the method cannot identify the form of the combinations. More recent studies have combined the FIM method with techniques for practical identifiability or subset selection for ordinary differential equation models to determine subsets of parameters that can be estimated from given data \cite{cintron2009sensitivity,eisenberg2014determining}. 

While FIM reflects local structural identifiability, one framework to assess generic structural identifiability is based on differential algebraic methods. This framework was initially developed for ordinary differential equations models but was recently extended in \cite{renardy2022structural} to age-structured PDE models for disease spread. This approach requires converting the model system to input-output equations consisting of a set of monic polynomial equations expressed in terms of the known model output $y$ and its derivatives, as well as in terms of rational coefficients depending on the model parameters $\vect{\theta}$ \cite{renardy2022structural}. This work builds on a substitution-based approach as in \cite{eisenberg2013identifiability,eisenberg2019input} to eliminate unobserved variables and to obtain a system whose identifiability features are equivalent to those of the original system. Specifically, identifiability is evaluated based on the coefficients of the monomial terms in the reduced system \cite{renardy2022structural}.

\subsection{Practical identifiability using Bayesian inference}\label{sec:MCMC_DRAM} 
A commonly used method for assessing practical identifiability is Bayesian inference using Markov Chain Monte Carlo (MCMC) sampling \cite{hines2014determination, simpson2020practical}. As described above, suppose we are interested in observed data $y$ and model parameter $\vect{\theta}$. Then, according to Bayes' theorem,
\begin{align}
    p(\vect{\theta} \vert y) \propto p(y \vert \vect{\theta}) p(\vect{\theta}),
\end{align}
where $p(\vect{\theta})$ denotes the prior distribution of $\vect{\theta}$, $p(y \vert \vect{\theta})$ denotes the likelihood function, and $p(\vect{\theta} \vert y)$ denotes the posterior distribution of $\vect{\theta}$. The likelihood function represents the extra information that $y$ contributes to our understanding of $\vect{\theta}$. In the Bayesian inference approach, we seek to estimate the posterior distribution, which specifies the distribution $\vect{\theta}$ given our knowledge of $y$ and $p(\vect{\theta})$. Here, we can consider $\vect{\theta}$ to be identifiable if we can estimate a relatively concentrated posterior.

We estimate the posterior distributions of parameters in the model using MCMC simulation. Specifically, we use the Delayed Rejection and Adaptive Metropolis (DRAM) MCMC algorithm, a variation of the Metropolis--Hastings MCMC algorithm \cite{haario2006dram}. A standard Metropolis--Hastings algorithm starts with a Markov Chain at initial position $\vect{\theta}_{i}$ and accepts candidate move $\vect{\theta}_{i+1}$ with probability $\alpha$, where
\begin{align}
    \alpha &= \min \biggr[1, \frac{p(\vect{\theta}_{i+1} \vert y)}{p(\vect{\theta}_{i} \vert y)}\biggr].
\end{align}
Proposals in Metropolis--Hastings are sampled from a multivariate normal distribution \cite{simpson2020practical}.

The DRAM algorithm has two advantages over the standard Metropolis--Hastings: DRAM incorporates (1) delayed rejection and (2) adaptive Metropolis samplers \cite{haario2006dram}. After the standard Metropolis--Hastings rejects a candidate move, delayed rejection proposes subsequent moves in lieu of remaining at the same position. With an adaptive Metropolis approach, the proposal distribution of Metropolis--Hastings is based on past samples in the Markov chain. Combined, adaptive Metropolis enhances DRAM's ability to explore the range of good proposal distributions, while delayed rejection improves DRAM's flexibility in its local exploration of the parameter space \cite{haario2006dram}.

Practical identifiability in a Bayesian setting can be determined graphically or through diagnostic statistics. In general, characteristics like poorly converging Markov Chains, label-switching, and multimodal or overly broad distributions indicate poor identifiability \cite{simpson2020practical,hines2014determination,siekmann2012mcmc}.

\subsection{Practical identifiability using profile likelihood analysis}\label{sec:profile_likelihood}

Computing the full MCMC posterior distributions for parameters of interest as described in Section~\ref{sec:MCMC_DRAM} is known to be computationally expensive, especially for partial differential equations models \cite{simpson2020practical}. An alternative approach to assessing practical parameter identifiability is to carry out a profile likelihood analysis \textblue{\cite{murphy2000profile}}. This method requires setting up the normalized likelihood function
\begin{align*}
    \mathcal{L}(\vect{\theta} ; y) &= \frac{p(\vect{\theta};y)}{\sup_{\vect{\vartheta}} p(\vect{\vartheta};y)}\,,
\end{align*}
where $p(\vect{\theta};y)$ is the likelihood function as in Section~\ref{sec:MCMC_DRAM}. This normalized likelihood assumes fixed data $y$ and is a function of the parameters $\vect{\theta}$. We then let $\vect{\theta}=(\psi, \vect{\lambda})$, where $\psi$ is a scalar parameter of interest whose identifiability we are interested in assessing, and $\vect{\lambda}$ is a vector of nuisance parameters. The profile likelihood for the interest parameter $\psi$ is then given by:
\begin{align}\label{eq:profile_likelihood}
    \mathcal{L}_p(\psi;y) &= \max_{\vect{\lambda}} \mathcal{L}(\psi, \vect{\lambda};y)\,.
\end{align}
In practice, this means that for each value of parameter $\psi$ chosen from a grid in an appropriate interval \textblue{around the nominal value}, parameters $\vect{\lambda}$ are optimized out. This yields optimal nuisance parameter values $\lambda^*(\psi)$ for each grid value of $\psi$; see \cite{simpson2020practical}. 

If the measurement noise is assumed to be normally distributed as $\epsilon \sim N(0,\sigma^2)$, then:
\begin{align}\label{eq:profile_likelihood_nsig}
    p(y;\psi,\vect{\lambda}) &= \left(\frac{1}{2 \pi \sigma^2}\right)^{n/2} \exp{ \left(-\frac{1}{2 \sigma^2} \|y-y_\mathrm{sim}(\psi,\vect{\lambda})\|^2\right)}\,,
\end{align}
where $y_\mathrm{sim}$ consists of the model solutions at $n$ time points \cite{raue2013joining,simpson2020practical}. The profiling calculation in equation~\eqref{eq:profile_likelihood} is then equivalent to solving a nonlinear least-squares optimization problem for each grid value of the parameter of interest $\psi$. 

\begin{figure}[t]
\center
\includegraphics[width=120mm]{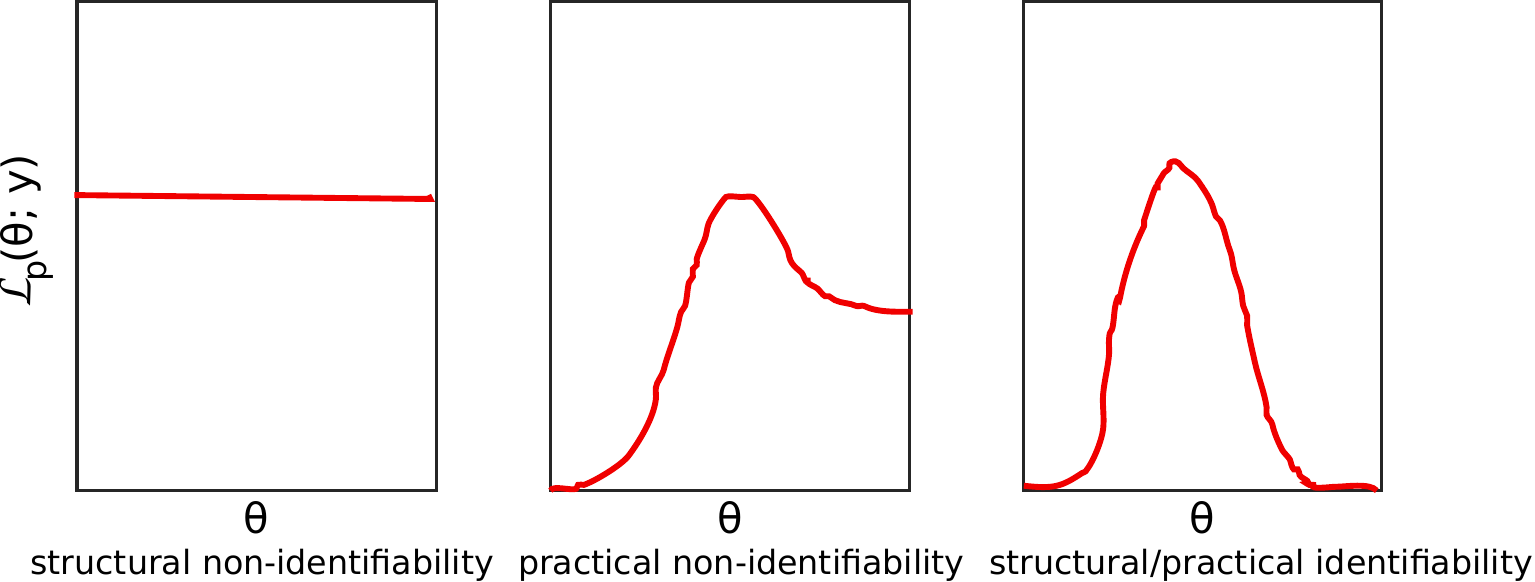}
\caption{Interpretation of profile likelihoods in terms of structural and practical identifiability \cite{raue2013joining}. \textblue{A flat likelihood (left) corresponds to structural non-identifiability, a profile that does not decrease to $0$ on one or both sides of the maximum (center) indicates practically non-identifiability, and a profile with a fast decrease to $0$ on both sides of the maximum (right) shows both structural and practical identifiability.}}
\label{fig:PL}
\end{figure}

The shape of the profile likelihoods can provide rich information about whether parameters can be inferred from measurement data \cite{raue2009structural,raue2013joining}, as can be seen in the cartoon in Figure~\ref{fig:PL} (inspired from \cite{raue2013joining}). \textblue{A completely flat profile as shown in the left panel of Figure~\ref{fig:PL} corresponds to a structurally non-identifiable parameter of interest, since the likelihood function does not change with parameter $\theta$. If the profile achieves a maximum but stays above a plateau value on one or both sides as parameter $\theta$ is varied, this indicates a practically non-identifiable parameter. Finally, a fast decrease of the profile to $0$ on both sides of the maximum shows both structural and practical parameter identifiability \cite{raue2009structural,raue2013joining}.} Since flat regions of the likelihood profile indicate that the parameter is practically or structurally unidentifiable \cite{eisenberg2014determining,raue2009structural,raue2013joining}, it is sometimes useful to examine the relationship between the interest parameter and each fitted nuisance parameter, in the flat regions of the likelihood \cite{eisenberg2013identifiability,eisenberg2014determining}. These are called subset profiles and can help uncover the form of potential identifiable combinations of parameters for the given model \cite{eisenberg2013identifiability,eisenberg2014determining}.

\section{Applications of practical and structural parameter identifiability to PDE models for synthetic FRAP data}\label{sec:app_identifiability}

We now turn to applying the parameter identifiability methods outlined in Section~\ref{sec:identifiability} for the PDE model~\eqref{eq:diff_pause} describing the dynamics of PTBP3 protein during fluorescence recovery after photobleaching. We consider \textblue{noiseless} synthetic FRAP datasets generated using the parameter regimes outlined in Table~\ref{tab:par_regimes} and determined based on the procedure in Section~\ref{sec:par_estimation}. In the notation of Section~\ref{sec:identifiability}, the relevant model output is $y(t;\vect{\theta}) = \mathrm{FRAP}(t;\vect{\theta)}$ as defined in equation~\eqref{eq:FRAP_curve}.

\subsection{Structural identifiability}\label{sec:app_structural}

First, we aim to determine the local structural identifiability of kinetic model parameters $D,\beta_1,\beta_2$ given completely accurate FRAP data using the Fisher information matrix method described in Section~\ref{sec:structural}. To construct this matrix, we first calculate the sensitivities of the output with respect to the model parameters. For example, we seek:
\begin{align*}
    \frac{\partial \mathrm{FRAP}}{\partial D}(t;\vect{\theta}) &= \int_{-l/2}^{l/2} \int_{-l/2}^{l/2} \left(f_D + c_D\right)(x,y,t;\vect{\theta}) \,\mathrm{d}x \,\mathrm{d}y   
\end{align*}
where $f_D = \frac{\partial f}{\partial D} $ and $c_D = \frac{\partial c}{\partial D}$. \textblue{By differentiating the PDE model system \eqref{eq:diff_pause} with respect to each parameter of interest, we obtain the following sensitivity equations for} the partial derivatives of the protein concentrations: 
\begin{align}
    \frac{\partial f_D}{\partial t} & = D \Delta f_D -\beta_2 f_D + \beta_1 c_D + \Delta f \,, \nonumber\\
    \frac{\partial c_D}{\partial t} & = -\beta_1 c_D + \beta_2 f_D \,, \nonumber\\
    \frac{\partial f_{\beta_1}}{\partial t} & = D \Delta f_{\beta_1} -\beta_2 f_{\beta_1} + \beta_1 c_{\beta_1} + c \,, \nonumber\\
    \frac{\partial c_{\beta_1}}{\partial t} & = \beta_2 f_{\beta_1} -\beta_1 c_{\beta_1}  -c \,, \nonumber\\
    \frac{\partial f_{\beta_2}}{\partial t} & = D \Delta f_{\beta_2} -\beta_2 f_{\beta_2} + \beta_1 c_{\beta_2} - f \,, \nonumber\\
    \frac{\partial c_{\beta_2}}{\partial t} & = \beta_2 f_{\beta_2} -\beta_1 c_{\beta_2} + f  \,.
\end{align}
\textblue{We provide additional details on the derivation of these equations in Appendix~\ref{sec:appendix_senseqs}.} We solve the above sensitivity equations simultaneously with integrating model~\eqref{eq:diff_pause} using the numerical methods outlined in Section~\ref{sec:par_estimation}. Then the sensitivity matrix is given by:
\begin{align}
    X &= \begin{pmatrix} \frac{\frac{\partial{\mathrm{FRAP}}}{\partial{D}}(\vect{t};\vect{\theta^0})}{\|\frac{\partial{\mathrm{FRAP}}}{\partial{D}}(\vect{t};\vect{\theta^0})\|_2} & \frac{\frac{\partial{\mathrm{FRAP}}}{\partial{\beta_1}}(\vect{t};\vect{\theta^0})}{\|\frac{\partial{\mathrm{FRAP}}}{\partial{\beta_1}}(\vect{t};\vect{\theta^0})\|_2} &  \frac{\frac{\partial{\mathrm{FRAP}}}{\partial{\beta_2}}(\vect{t};\vect{\theta^0})}{\|\frac{\partial{\mathrm{FRAP}}}{\partial{\beta_2}}(\vect{t};\vect{\theta^0})\|_2}  \end{pmatrix}\,,
\end{align}
where $\vect{\theta^0}$ corresponds to the baseline parameter regimes of interest in Table~\ref{tab:par_regimes}. Here we have normalized each column by the $L^2$ norm of the corresponding sensitivity vector, to account for the different magnitudes of the parameters. 

The Fisher Information Matrix $F = X^T X$ is then a $3 \times 3$ matrix whose determinant \textblue{and eigenvalues are displayed in the last two columns} of Table~\ref{tab:par_regimes} for the relevant parameter regimes considered. As expected, the matrix is singular for the effective diffusion parameter regime (Set 4), where we only expect to be able to identify one parameter combination (effective diffusion): \textblue{the expectation that we can identify only a single parameter combination is also reflected by the observation that two of the three eigenvalues of $F$ vanish.} For the other parameter regimes corresponding to wild-type and mutant protein binding settings (Sets 1-3), the determinant of the matrix is small, however it is difficult to conclude whether all or only subsets of the parameters are locally structurally identifiable given FRAP recovery data. 

To assess general structural identifiability, we also apply the differential algebra approach recently outlined in \cite{renardy2022structural} for age-dependent PDE models. We focus on the simplification of the reaction-diffusion PDE model \eqref{eq:diff_pause} to one spatial dimension $x$. Since FRAP recovery data requires averaging out the sum of the protein concentrations in each state over a given spatial domain corresponding to the bleaching region ($x\in [-l/2,l/2]$), we start by considering model output:
\begin{align*}
    z(x,t) &= f(x,t) + c(x,t)\,.
\end{align*}
Re-writing system~\eqref{eq:diff_pause} in terms of the total protein concentration $z(x,t)$ and the concentration of bound protein complexes $c(x,t)$ yields:
\begin{align}\label{eq:diff_pause_z}
    \frac{\partial z}{\partial t} &= D z_{xx} - D c_{xx}\,, \nonumber \\
    \frac{\partial c}{\partial t} &= \beta_2 z - (\beta_1+\beta_2)c\,.
\end{align}

The goal is to express this system in terms of model output $z$ and its derivatives. By differentiating the first equation in \eqref{eq:diff_pause_z} with respect to time and using substitution to eliminate variable $c$, we obtain:
\begin{align}\label{eq:z_input_output}
    0 &= -z_{tt}+ D z_{txx} + \beta_1D z_{xx} - (\beta_1+\beta_2)z_t\,.
\end{align}
This input-output equation is written as a polynomial equation in terms of derivatives of $z$. As in \cite{renardy2022structural}, we rank the terms within the polynomial by assuming that derivatives with respect to time are ranked higher than those with respect to space. To ensure a monic polynomial in $z_{txx}$, we therefore divide equation~\eqref{eq:z_input_output} by $D$ and obtain the following set of polynomial coefficients:
$\{1, -\frac{1}{D}, \beta_1, -\frac{\beta_1+\beta_2}{D} \}$. This provides a map from parameter space to the polynomial coefficients, which can be used to determine identifiability information for the model equations \cite{renardy2022structural}. \textblue{In particular, this set of polynomial coefficients is interpreted as a function $\phi(D,\beta_1,\beta_2)$ of the model parameters so that, in our case, we have $\phi(D,\beta_1,\beta_2):=(1, -\frac{1}{D}, \beta_1, -\frac{\beta_1+\beta_2}{D})$. If this map is injective, then the model is structurally identifiable. In our case, $\phi(D,\beta_1,\beta_2)$ is indeed injective, which} suggests that the parameters $\{D,\beta_1,\beta_2\}$ are structurally identifiable, provided that the time and spatial derivatives of $z$ in equation~\eqref{eq:z_input_output} are available. \textblue{This is consistent with recent results on structural identifiability of reaction-diffusion models in \cite{browning2023structural}.}

In practice, however, the total protein fluorescence concentration through time and space $z(t,x)$ is often not available from fluorescence recovery experiments or is only accessible as very noisy and diffuse images. In the rare instances when this is available, derivatives of this concentration would need to be numerically approximated, incurring additional errors. Typically, the only available measurement data from FRAP experiments is the spatially-averaged quantity $y(t) = \int_{-l/2}^{l/2} z(x,t) \,\mathrm{d}x$ (or equation~\eqref{eq:FRAP_curve} for the 2-dimensional system), which provides substantially less information. \textblue{As mentioned in Section~\ref{sec:bio_motivation}, this is due to the low spatial resolution of the FRAP experiments and to their focus on providing insights into the timing of dynamic protein processes \cite{ishikawa2012advanced}.} For $y(t)$, Equation~\eqref{eq:z_input_output} becomes:
\begin{align}\label{eq:y_input_output}
    0 &= -y_{tt}+ 2D z_{tx}(l/2,t) + 2 \beta_1 D z_{x}(l/2,t) - (\beta_1+\beta_2)y_t\,.
\end{align}
The derivatives of the total concentration of protein at the boundaries of the bleach point are however not available from the data. Therefore, the current framework for using the differential algebraic approach cannot provide insight into structural identifiability of the model parameters in this setting.

\subsection{Practical identifiability using Bayesian inference}\label{sec:app_MCMC_DRAM}

We then investigate the practical identifiability of parameters $D, \beta_1, \beta_2$ given FRAP data using the MCMC DRAM algorithm described in Section~\ref{sec:MCMC_DRAM}. In addition, we use this Bayesian inference approach to study the practical identifiability of parameters $p$ (from equation~\eqref{eq:IC_final}) and $C_{0}$ (from equation~\eqref{eq:Cb_full_approx}). We start with initial parameter guesses $D^{*}, \beta_1^{*}, \beta_2^{*}$, $p^{*}$, $C_{0}^{*}$ determined as outlined in Section~\ref{sec:par_estimation}. Since MCMC DRAM requires its sampling intervals to be bounded \cite{haario2006dram}, we set parameter bounds that are one order of magnitude lower and higher than the initial guesses. The exceptions are $p$, where the maximum bound is $1$ (since it denotes a fraction), and $C_0$, where the maximum bound is set to $C_{0}^{*} + 1$.

We carry out MCMC DRAM on synthetic FRAP data generated using the parameter sets in Table~\ref{tab:par_regimes} for $10,000$ sampling iterations. We determine convergence of the resulting Markov Chains using the Geweke diagnostic test \cite{geweke1991evaluating}. A higher Geweke test score indicates a higher probability of convergence in the corresponding Markov Chain. Table~\ref{tab:geweke} shows that, while the Geweke test suggests strong convergence in the Markov Chains at $10,000$ iterations for $D$, $p$, $C_{0}$ and moderate convergence for $\beta_{1}$, there does not appear to be strong evidence of convergence for $\beta_{2}$, despite the large number of iterations. 

\begin{figure}[t]
\center
\includegraphics[width=120mm]{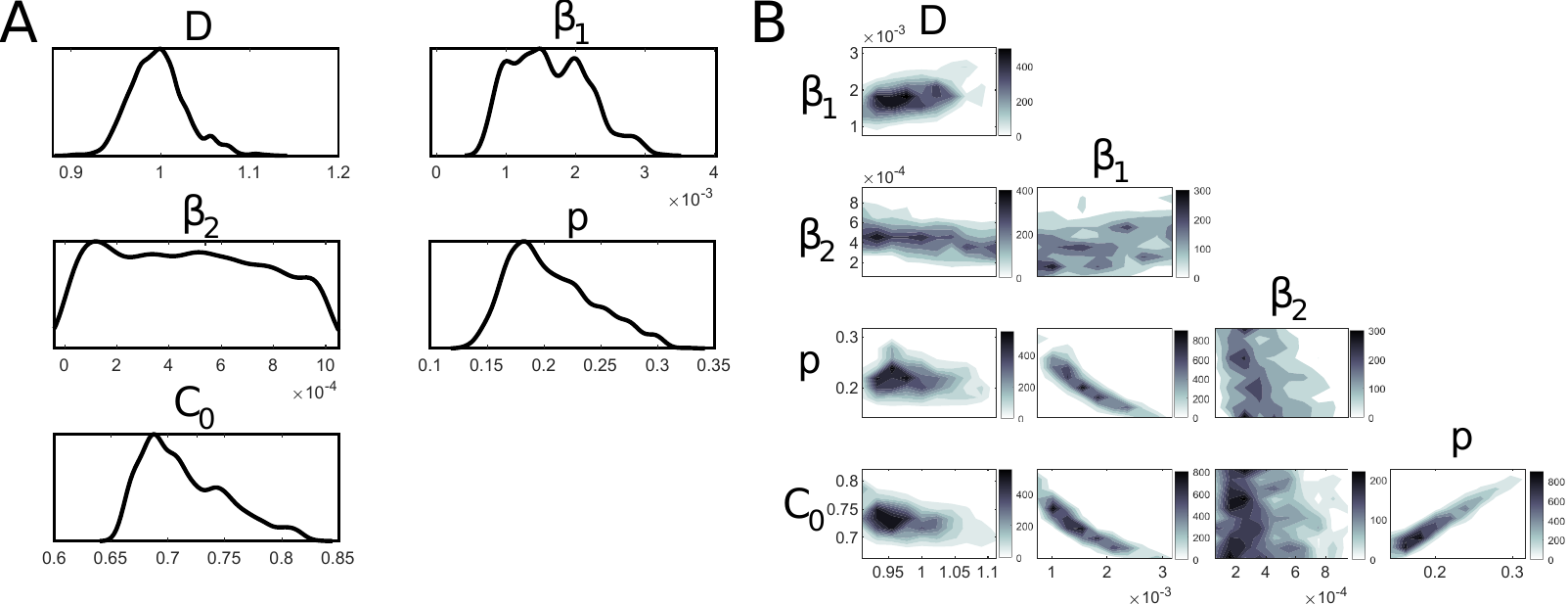}
\caption{MCMC DRAM-estimated A) univariate and B) bivariate marginal distributions for noiseless FRAP data generated using Parameter Set 2 in Table~\ref{tab:par_regimes}. \textblue{Scale bars in Panel B correspond to the number of sampled points in the MCMC simulation for each parameter pair.}}
\label{fig:mcmc_dram}
\end{figure}

To determine practical identifiability based on MCMC DRAM, we study the univariate and bivariate marginal parameter distributions estimated by the inference algorithm. Across the Table~\ref{tab:par_regimes} parameter regimes, we find that some of the MCMC DRAM-estimated marginal distributions for $D$, $\beta_{1}$, $\beta_{2}$, $p$, and $C_{0}$ appear broad or multimodal, suggesting a lack of practical identifiability. Figure~\ref{fig:mcmc_dram} shows the estimated parameter distributions for Parameter Set 2, where the distribution of rate $\beta_2$ is especially broad. 

In addition, assessing practical identifiability using this Metropolis–Hastings MCMC algorithm carries a high computational cost. Study \cite{simpson2020practical} also observed this for applications to PDE models of cell scratch assays. We find that the method is even less computationally feasible for the FRAP model, where the concentrations of interest are tracked in two spatial dimensions. \textblue{In particular, parameter distributions estimated using MCMC DRAM as shown in Figure~\ref{fig:mcmc_dram} take 18--22 hours each to simulate on a standard computer cluster. }

\begin{table}[ht]
\begin{center}
\caption{Geweke Diagnostic Scores for parameter convergence using MCMC DRAM carried out for the parameter sets in Table~\ref{tab:par_regimes}.}\label{tab:geweke}
\begin{tabular}{@{}llllll@{}}
\toprule
Parameters & $D$  & $\beta_1$ & $\beta_2$ & $p$ & $C_0$\\
\midrule
Set 1    & $0.878$   & $0.639$  & $0.492$  & $0.838$ & $0.997$\\
Set 2    & $0.996$   & $0.414$  & $0.129$  & $0.741$ & $0.931$\\
Set 3    & $0.975$   & $0.602$  & $0.573$  & $0.961$ & $0.936$\\
\end{tabular}
\end{center}
\end{table}

\subsection{Practical identifiability using profile likelihood analysis}\label{sec:app_profile_likelihood}

We next compute profile likelihoods for the kinetic parameters of interest ($D,\beta_1,\beta_2$) in the FRAP model. By visualizing the residuals from fitting the experimental FRAP data using model~\eqref{eq:diff_pause} as described in Section~\ref{sec:par_estimation}, we conclude that the observation noise can be assumed to be normally distributed for the purpose of our application. We therefore choose the fixed standard deviation of the measurement noise in equation~\eqref{eq:profile_likelihood_nsig} as $\sigma = 0.1 * \mathrm{mean}(\mathrm{FRAP}_\mathrm{gen}(t))$ based on the true synthetically-generated FRAP curve corresponding to each wild-type or mutant parameter regime. The profile likelihood calculation then reduces to carrying out nonlinear least-squares optimization to optimize out the nuisance parameters, which we carry out using the \texttt{lsqnonlin} function in Matlab. 

\begin{figure}[t]
\center
\includegraphics[width=100mm]{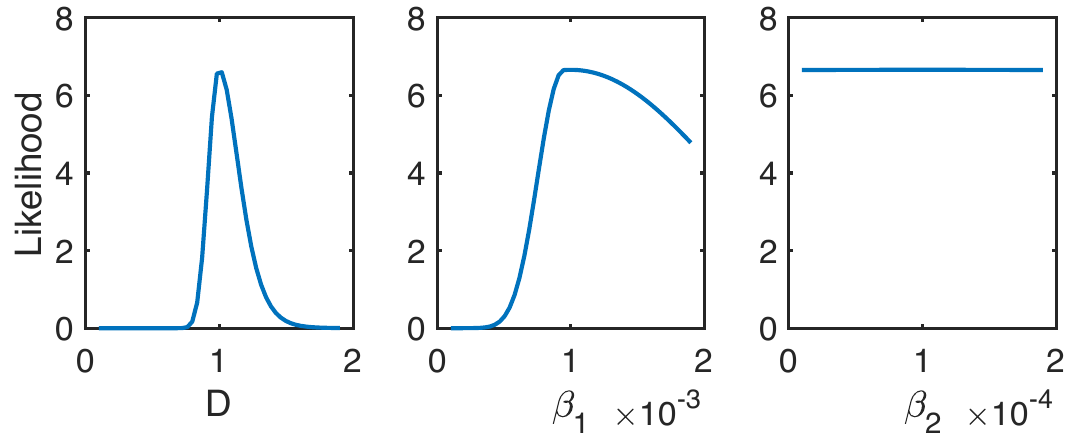}
\caption{Profile likelihoods for each interest parameter on the $x$ axis given noiseless FRAP data synthetically generated using model~\eqref{eq:diff_pause} and Parameter Set 2 in Table~\ref{tab:par_regimes}.}
\label{fig:prof_likelihood}
\end{figure}

For example, recall from Section~\ref{sec:profile_likelihood} that, when interested in the identifiability of the diffusion coefficient $D$, we fix values of $D$ from an appropriate grid. We use a uniform grid for parameter $D$ on an interval given by $[D^{*}/10, 10 D^{*}]$, where $D^{*}$ is the starting parameter guess determined through the initial deterministic procedure outlined in Section~\ref{sec:par_estimation}. For each value of $D$ in this grid, we maximize the profile likelihood (equation~\eqref{eq:profile_likelihood}), which yields values $\beta_1^*(D)$ and $\beta_2^*(D)$ for the optimized nuisance parameters. The likelihood of each parameter of interest is visualized in Figure~\ref{fig:prof_likelihood} for FRAP data generated using Parameter Set 2. While \textblue{$D$ appears to be identifiable given our model and for this generated dataset, $\beta_1$ and $\beta_2$ are both practically non-identifiable, even with perfect synthetic FRAP data. Similar results, where the rates $\beta_1$ and $\beta_2$ are both practically unidentifiable, are observed in the other parameter regimes. Therefore, profile likelihood analysis suggests that the model switching rates are practically unidentifiable given the information typically captured in FRAP data. This is consistent with the relatively small determinants of the Fisher Information Matrices for assessing structural identifiability of parameters that we listed in Table~\ref{tab:par_regimes} for all parameter sets (see Section~\ref{sec:app_structural}). We summarize this step in flowchart Figure~\ref{fig:flowchart}C.}

\begin{figure}[t]
\center
\includegraphics[width=90mm]{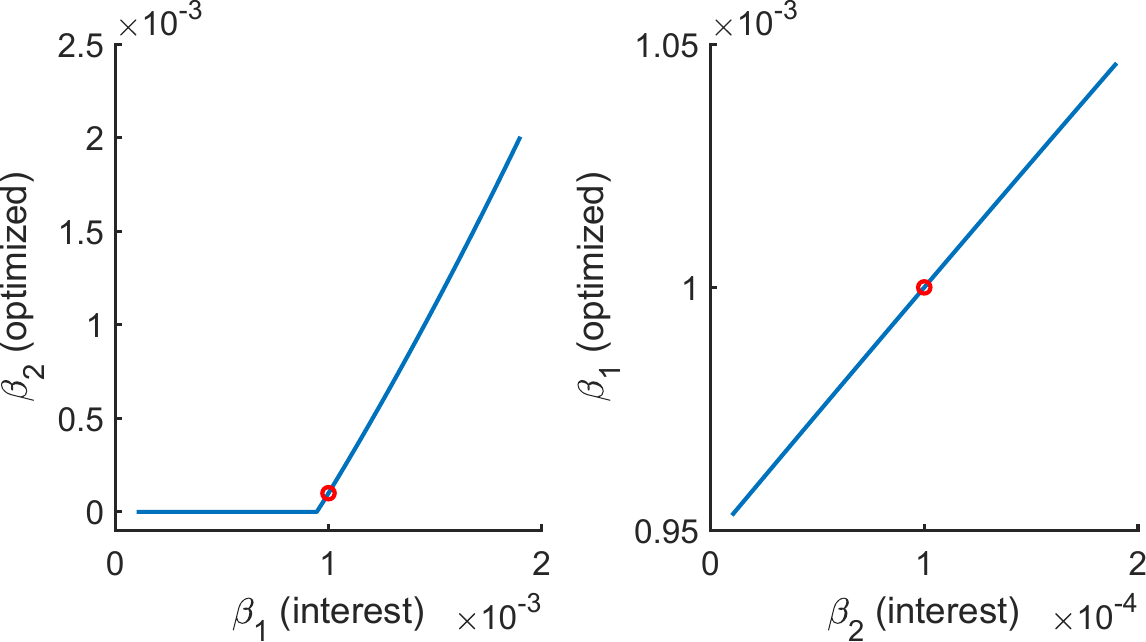}
\caption{Subset profiles for each interest rate parameter on the $x$ axis and the corresponding optimized nuisance rate parameter on the $y$ axis given noiseless FRAP data synthetically generated using model~\eqref{eq:diff_pause} and Parameter Set 2 in Table~\ref{tab:par_regimes}. The true reaction rate parameters are indicated with red circles. }
\label{fig:subset_profile}
\end{figure}

Profile likelihood analysis also provides all the information needed to generate subset profiles, which in this case help visualize the relationship between each rate as an interest parameter and the other rate as the optimized nuisance parameter, following the approach in \cite{eisenberg2013identifiability}. Figure~\ref{fig:subset_profile} shows the inferred linear relationship between the rate parameters. Since we explore the application of the methods to synthetically-generated FRAP recovery curves, the true values of the parameters are indicated using a red circle in Figure~\ref{fig:subset_profile}. We find that the true parameters indeed lie on the curves outlining the relationship between the reaction rates. 

\section{Investigating parameter relationships in FRAP models} \label{sec:par_combinations}

\textblue{The investigation of established methods of parameter identifiability in Section~\ref{sec:app_identifiability} shows that the current strategies do not provide insight into structural identifiability of the model parameters and that some of the reaction rates may be practically unidentifiable. In this Section, we formulate and implement a methodology for investigating practical parameter identifiability in PDE models of FRAP experiments that extends the approach proposed in \cite[Section~3]{eisenberg2014determining} from ODE models to both PDE models and experimental data. The framework we present here builds on the calculation of profile likelihoods for these model parameters (see Sections~\ref{sec:profile_likelihood} and \ref{sec:app_profile_likelihood}), which is not computationally expensive, and allows us to determine which parameters are practically identifiable as well as to identify parameter combinations that can be inferred given FRAP experimental data.}

\begin{figure}[t]
\center
\includegraphics[width=100mm]{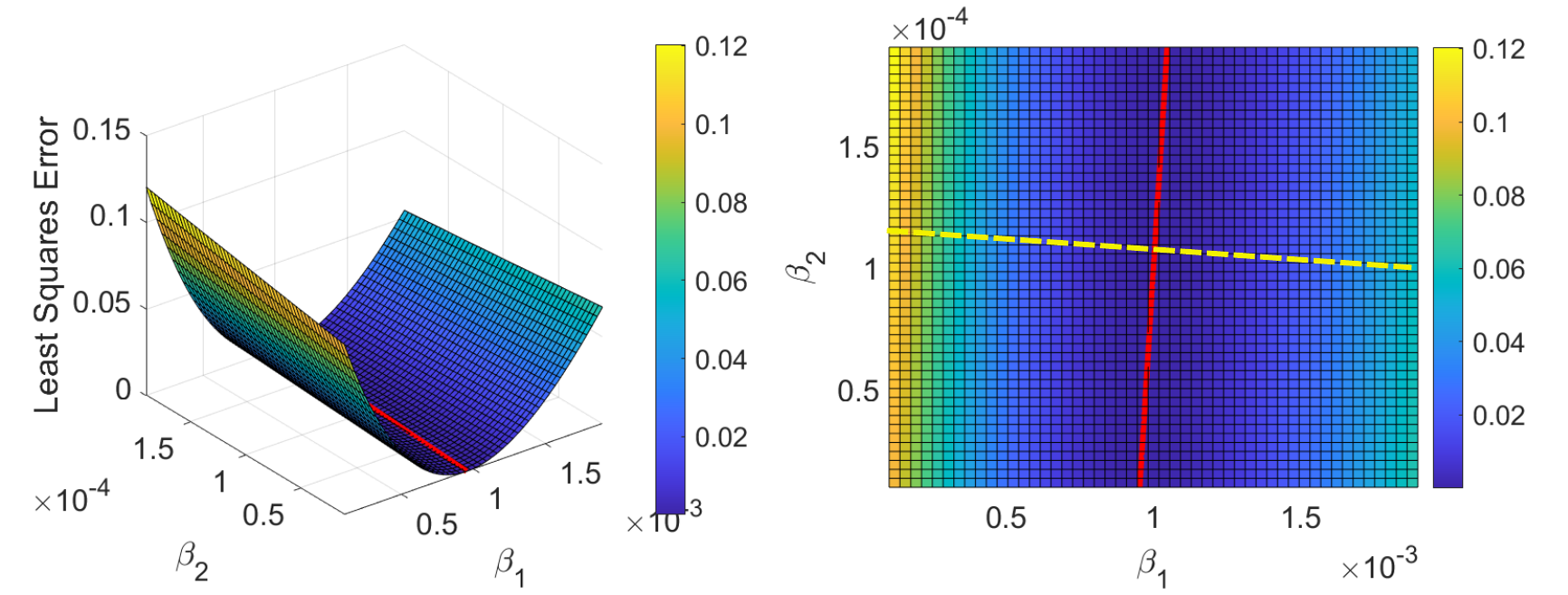}
\caption{Contour plots of the least-squares error between FRAP data generated using $D=1\mu m^2/s$ and rates $\beta_1$ and $\beta_2$ from the grid shown and ground-truth synthetic data generated using Parameter Set 2 in Table~\ref{tab:par_regimes}. \textblue{The likelihood is minimized along the entire red curve.}}
\label{fig:contour_plot}
\end{figure}

\textblue{To motivate our algorithm, we recall that the reaction rates in our model are consistently unidentifiable based on the Bayesian inference and profile likelihood methods described in Section~\ref{sec:app_identifiability}, while the diffusion constant $D$ is identifiable. Our goal is therefore to explore the likelihood landscape by varying the reaction rates $\beta_1$ and $\beta_2$ in a parameter grid for a fixed value of $D$. To do so, we set our ground-truth parameters to the values from Parameter Set~2 and compute a ground-truth FRAP curve by integrating model~\eqref{eq:diff_pause} for these parameter values. Next, we compute the least-squares error between the ground-truth FRAP curve and the FRAP curves from model~\eqref{eq:diff_pause} where we vary the reaction rates $\beta_1$ and $\beta_1$ on a square grid with $50$ equally spaced values between $10^{-4}/s$ and $2 \times 10^{-3}/s$ and between $10^{-5}/s$ and $2 \times 10^{-4}/s$, respectively, and a fixed value of $D=1\mu m^2/s$. Figure~\ref{fig:contour_plot} shows the contour plot of this error as a function of $(\beta_1,\beta_2)$: as is visible there, the likelihood is minimized along an entire curve (highlighted in red in Figure~\ref{fig:contour_plot}), and we therefore cannot distinguish points along this curve. This curve also coincides with the relationship between the optimized and interest parameter rates in the profile likelihood analysis in Figure~\ref{fig:subset_profile}. While contour plots as in Figure~\ref{fig:contour_plot} are different for each diffusion coefficient $D$, we observe similar behavior for the diffusion coefficients characterizing the other parameter regimes in Table~\ref{tab:par_regimes}.}

\begin{figure}[t]
\center
\includegraphics[width=80mm]{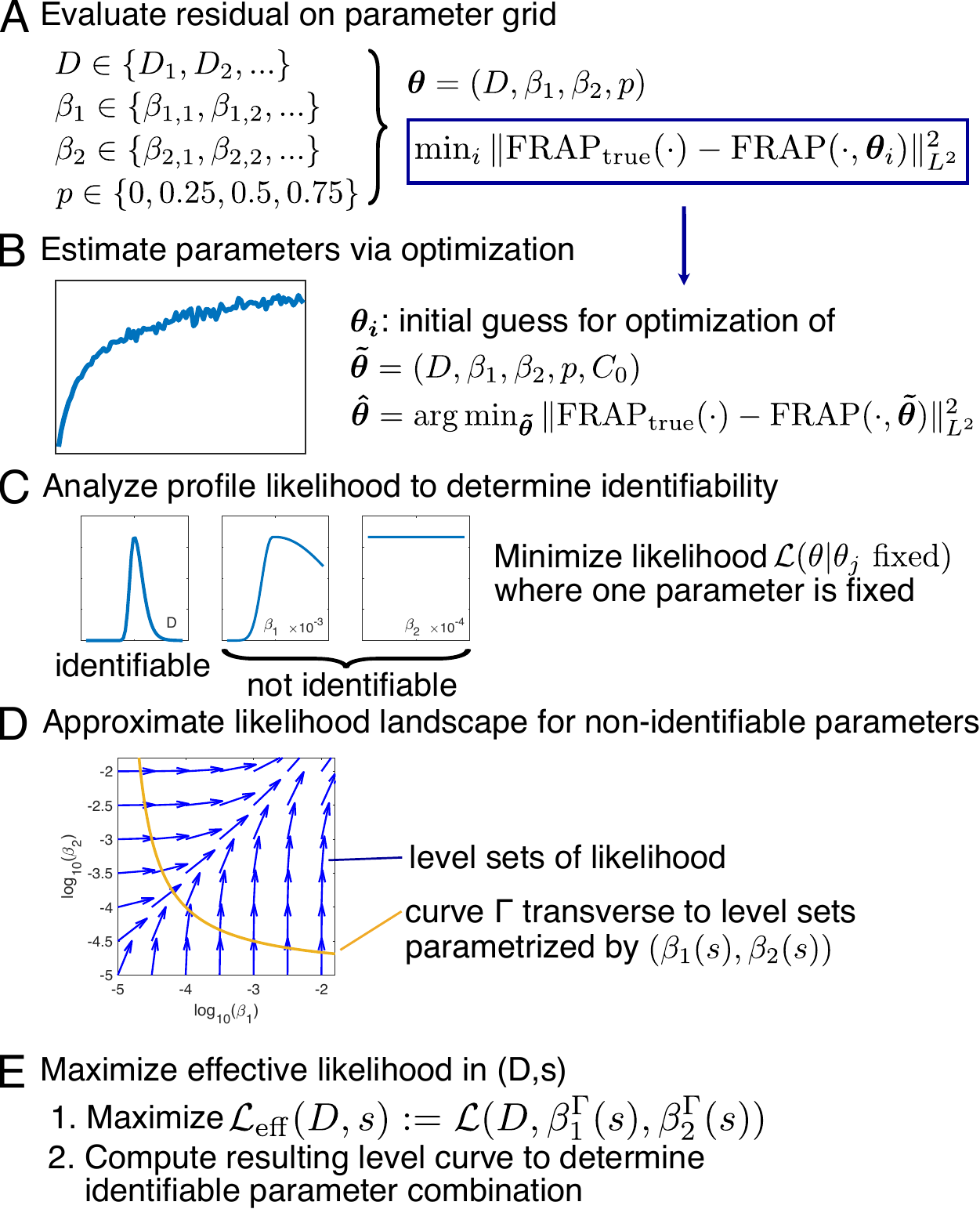}
\caption{This diagram summarizes our algorithm in a flowchart.}
\label{fig:flowchart}
\end{figure}

\textblue{We now turn these observations into an algorithm and refer to Figure~\ref{fig:flowchart} for a flowchart. Our goal is to (1) calculate a base point $Q^*$ on the minimizing curve as a regular minimum of an appropriate likelihood function and (2) use the base point $Q^*$ to compute the entire minimizing curve and therefore the identifiable combination of the reaction parameters. To achieve (1), we (i) select a curve $\Gamma$ in the $(\beta_1,\beta_2)$-plane transverse to the red curve, (ii) parametrize this curve in the form $(\beta_1^\Gamma,\beta_2^\Gamma)(s)$ by a parameter $s$ so that $(\beta_1^\Gamma,\beta_2^\Gamma)(s)$ traces out $\Gamma$ as $s$ varies, and (iii) minimize the likelihood function $\mathcal{L}(D,\beta_1^\Gamma(s),\beta_2^\Gamma(s); y)$ over $(D,s)$, where $y$ is the ground-truth FRAP curve. This optimization problem has a non-degenerate minimum $(D^*,s^*)$, since the parameters $(D,s)$ are identifiable. In particular, we obtain $Q^*=(D^*,\beta_1^\Gamma(s^*),\beta_2^\Gamma(s^*))$. To accomplish (2), we fix $D=D^*$ and use linear interpolation and a forward Euler scheme to compute the contour curve of the error function that passes through $Q^*$, which then provides the curve on which the reaction parameters $(\beta_1,\beta_2)$ must lie.}

\begin{figure}[t]
\center
\includegraphics[width=80mm]{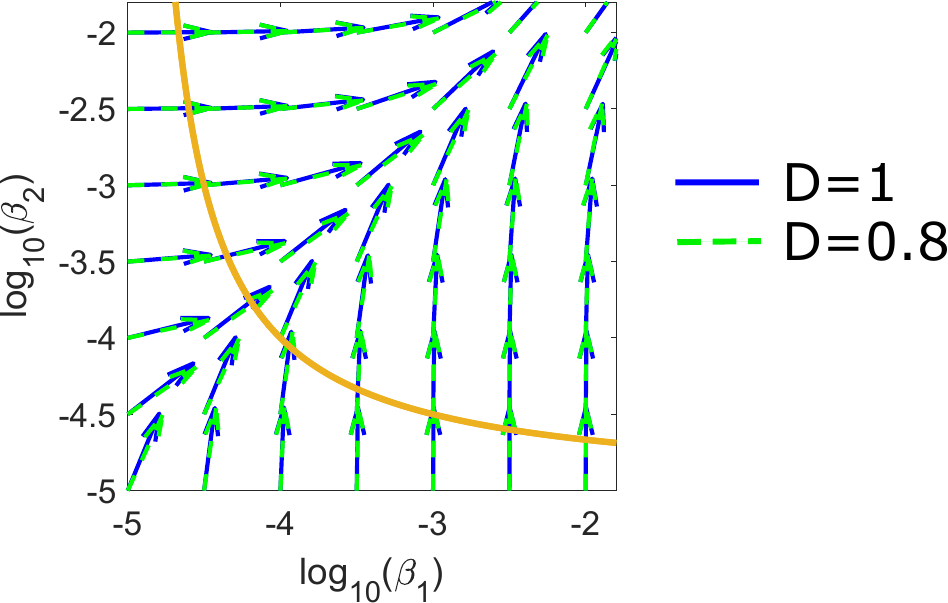}
\caption{Grid of inferred slopes based on profile likelihood analysis for the relationship between parameters $\beta_1$ and $\beta_2$ for FRAP datasets generated using $D=1\mu m^2/s$ (blue) or $D=0.8\mu m^2/s$ (green dashed) and the indicated reaction rates. The curve $\Gamma$ that
crosses the contour curves of the error function for $D=1\mu m^2/s$ transversely is shown in yellow. }
\label{fig:slope_grid}
\end{figure}

\textblue{Next, we discuss in more detail how we implement the proposed parameter-estimation algorithm. Since the diffusion constant $D$ is identifiable, we set it to the value found in the profile likelihood analysis. We then select a grid in the $(\log_{10}\beta_1,\log_{10}\beta_2)$-plane in order to inform our choice of the curve $\Gamma$. For each point on the grid (and the same fixed value of $D$), we generate synthetic FRAP recovery datasets from model~\eqref{eq:diff_pause} and calculate the resulting errors (that is, the $L^2$-differences between ground-truth and generated FRAP curves) on the parameter grid. We then use linear interpolation to compute the tangent vectors to the contour curves of the error function at the grid points. An example of the resulting vector field is shown in Figure~\ref{fig:slope_grid} (blue arrows), where we fixed the diffusion coefficient $D=1\mu m^2/s$ and selected seven values equally spaced on a log scale from $10^{-5}$ to $10^{-2}$ for the reaction rates $\beta_1$ and $\beta_2$. We can now choose a curve $\Gamma$ that crosses the contour curves of the error function transversely: we can either choose the transverse curve $\Gamma$ in explicit analytical form or else again use linear interpolation and a forward Euler scheme applied to the gradients of the vector field to compute such a transverse curve $\Gamma$ numerically. For illustration, we use the explicit analytical parametrization
\begin{align} \label{eq:gamma_curve}
\log_{10} \beta_1 &= s + \sqrt{s^2+1} - 5 \\
\log_{10} \beta_2 &= -s + \sqrt{s^2+1} - 5 \nonumber \,,
\end{align}
which yields the yellow curve $\Gamma$ shown in Figure~\ref{fig:slope_grid}. The assumption we make is that the chosen curve $\Gamma$ intersects each level curve transversely in a unique point for all nearby values of $D$. Figure~\ref{fig:slope_grid} indicates that the vector field generated with $D=0.8\mu m^2/s$ (green dashed arrows) is very similar to the one generated with $D=1\mu m^2/s$ (blue arrows), and that the curve $\Gamma$ chosen above (yellow) is still appropriate for this different diffusion coefficient. We summarize this step in flowchart Figure~\ref{fig:flowchart}D.}

\begin{figure}[t]
\center
\includegraphics[width=\textwidth]{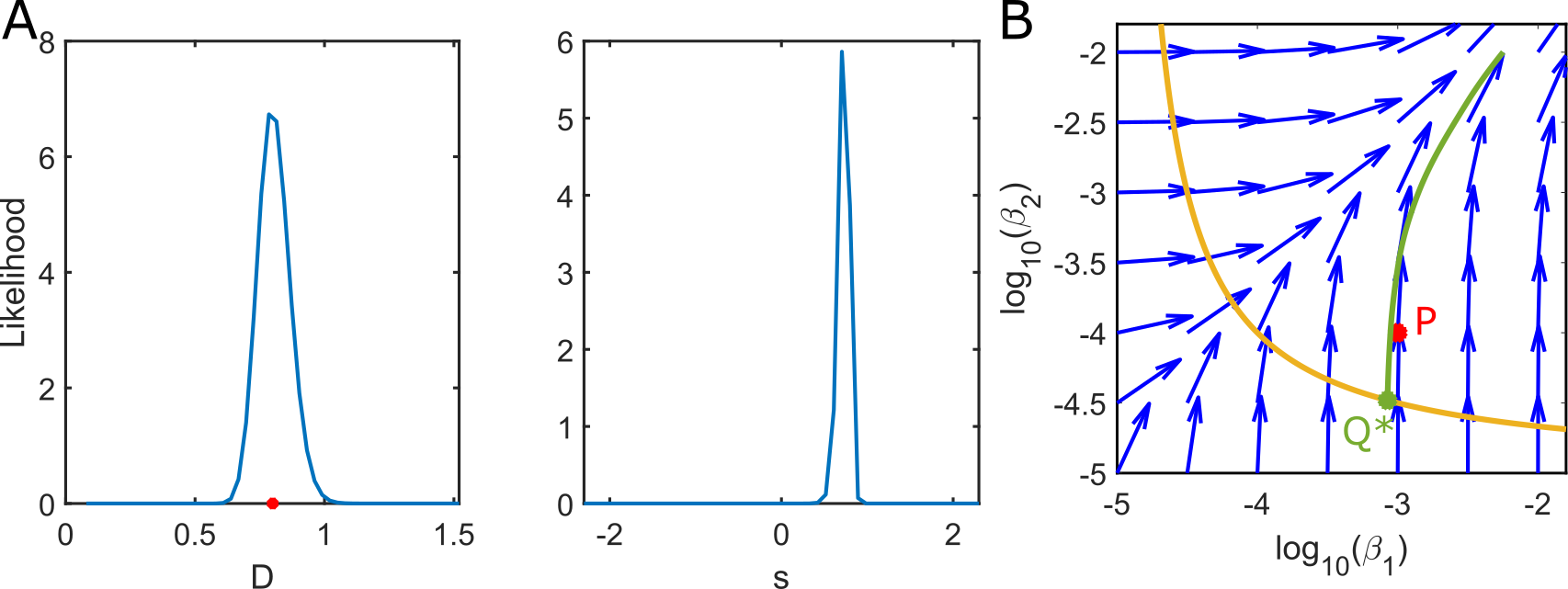}
\caption{(A) Profile likelihoods for each interest parameter on the $x$ axis (diffusion coefficient $D$ and parameter $s$ on curve $\Gamma$) given noiseless FRAP data generated using $D=0.8\mu m^2/s$ and rates as given in Parameter Set 2 in Table~\ref{tab:par_regimes}. The red star in the left panel corresponds to the ground-truth value $D_0=0.8\mu m^2/s$, while the maximum of the profile likelihood is achieved at $D^*=0.785\mu m^2/s$. (B) Grid of inferred slopes as in Figure~\ref{fig:slope_grid} (blue), overlaid with transverse curve $\Gamma$ (yellow), original ground-truth parameter set $P$ (red), and trace of error-minimizing contour curve as well as its intersection point $Q^*$ with the transverse curve (green). }
\label{fig:PL_s}
\end{figure}

\textblue{To demonstrate this proposed framework, we consider a FRAP dataset generated using $D^0=0.8\mu m^2/s$ and the reaction rates $\beta_1^0=10^{-3}$/s and $\beta_2^0=10^{-4}$/s from Parameter Set 2 to generate the ground-truth point $P$ shown in Figure~\ref{fig:PL_s}B. We use the parametrization of $\Gamma$ described in equation~\eqref{eq:gamma_curve}. Figure~\ref{fig:PL_s}A shows the results of the profile likelihood analysis from Section~\ref{sec:app_profile_likelihood} applied to the parameters ($D,s)$: the clear peaks in the profiles for these parameters demonstrate that the diffusion coefficient and the re-parametrized parameter $s$ are practically identifiable. The peak in the diffusion coefficient profile is achieved at $D^*=0.785 \mu m^2/s$, close to its true value of $D=0.8 \mu m^2/s$, while the peak in the $s$ profile is achieved at value $s^*=0.704$, which corresponds to the intersection point $Q^*=(\log_{10}\beta_1,\log_{10}\beta_2)$ of the contour curve (green star in Figure~\ref{fig:PL_s}B) along which the error function is minimized with the transverse curve $\Gamma$. Alternatively, we could have also optimized the likelihood function $\mathcal{L}(D,\beta_1^\Gamma(s),\beta_2^\Gamma(s); y)$ as described above. Focusing on the identified value of the diffusion coefficient $D^*$ (roughly $0.8 \mu m^2/s$), we generate contour plots as in Figure~\ref{fig:contour_plot} and slope grids as in Figure~\ref{fig:slope_grid}. Using linear interpolation and a forward Euler scheme for the tangent vector, we then numerically compute the contour curve of the error function that passes through $Q^*$, which then provides the curve on which the ground-truth parameters $(\log_{10}\beta_1,\log_{10}\beta_2)$ must lie (green curve in Figure~\ref{fig:PL_s}B). Notably, the ground-truth point $P$ (red star in Figure~\ref{fig:PL_s}B) is very close to this curve.}

\section{Application to an experimental FRAP dataset} \label{sec:app_experiment}

\begin{figure}[t]
\center
\includegraphics[width=\textwidth]{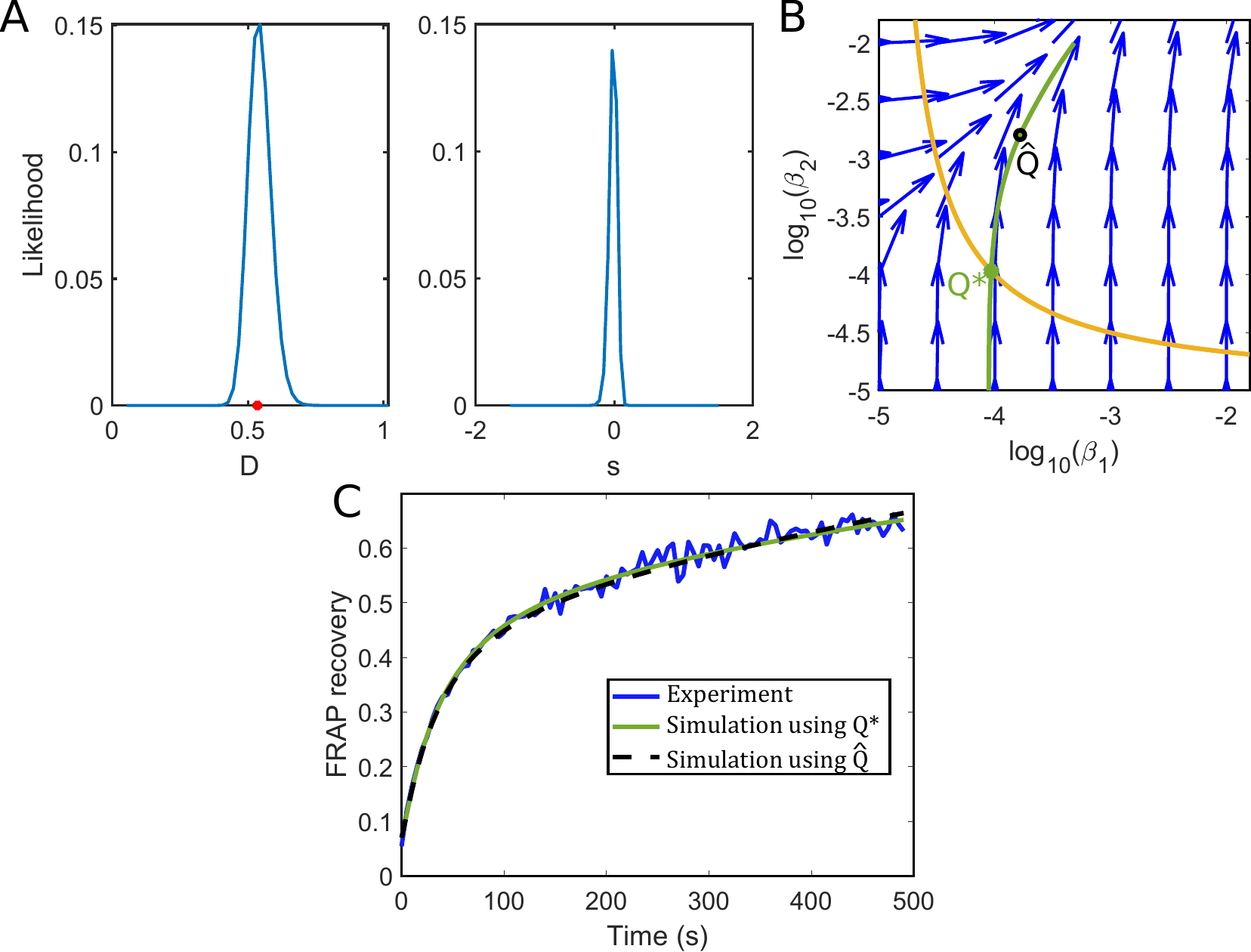}
\caption{(A) Profile likelihoods for each interest parameter on the $x$ axis (diffusion coefficient $D$ and parameter $s$) for a single RRM mutant experimental FRAP dataset. The red star in the left panel corresponds to the estimated value $D_0=0.535\mu m^2/s$, while the maximum of the profile likelihood is achieved at $D^*=0.545\mu m^2/s$. (B) Grid of inferred slopes for fixed diffusion coefficient $D_0=0.535\mu m^2/s$, overlaid with transverse curve $\Gamma$ (yellow), and trace of error-minimizing contour curve. The intersection point $Q^*$ of the error-minimizing curve with $\Gamma$ is denoted by a green star, while another point $\hat{Q}$ on the error-minimizing curve is shown as a black circle. (C) Fit of the experimental FRAP curve (blue) with simulated FRAP data generated using rate parameter sets given by $Q^*$ (green solid line) and by $\hat{Q}$ (black dashed line) indicated in panel (B).}
\label{fig:PL_s_mut3_expt}
\end{figure}

We also illustrate the application of the framework proposed in Section~\ref{sec:par_combinations} to a FRAP experimental dataset corresponding to the dynamics of PTBP3 protein with a single RRM mutant (mut3 in \cite{cabral2022multivalent}). This mutant has only one RNA-binding domain that can bind to the non-dynamic L-body RNA. We first carry out deterministic parameter estimation (as described in Section~\ref{sec:par_estimation}) for this fluorescence recovery dataset and obtain an estimate of the value of the diffusion coefficient $D_0\approx 0.535 \mu m^2/s$. We then fix this value for $D$ and vary the rate parameters on a grid in the $(\log_{10}\beta_1,\log_{10}\beta_2)$-plane to generate synthetic datasets and inform the choice of the transverse curve $\Gamma$. Figure~\ref{fig:PL_s_mut3_expt}B shows that the same choice of curve $\Gamma$ from equations~\eqref{eq:gamma_curve} is appropriate here as well.

As in the synthetic data setting investigated in Section~\ref{sec:par_combinations} and Figure~\ref{fig:PL_s}, we carry out profile likelihood analysis for parameters $D$ and $s$ for this experimental dataset. Figure~\ref{fig:PL_s_mut3_expt}A shows that the profiles for these parameters have clear peaks, indicating that they are practically identifiable. The peak in the diffusion coefficient profile is achieved at $D^*=0.545 \mu m^2/s$, close to the value we originally estimated. The peak in the $s$ profile is achieved at value $s^*=-0.031$. This value of $s^*$ identifies the intersection point $Q^*=(\log_{10}\beta_1,\log_{10}\beta_2)$ of the contour curve (green star in Figure~\ref{fig:PL_s_mut3_expt}B) along which the error function is minimized with the transverse curve $\Gamma$. We then use linear interpolation to numerically compute the curve on which we predict that the true parameters $(\log_{10}\beta_1,\log_{10}\beta_2)$ must lie on (green curve in Figure~\ref{fig:PL_s_mut3_expt}B). Panel C of Figure~\ref{fig:PL_s_mut3_expt} shows the original FRAP fluorescence recovery data (in blue) as well as the fit using two parameter sets chosen along the green curve in Figure~\ref{fig:PL_s_mut3_expt}B: $Q^*$ yields the green solid line curve fit in Figure~\ref{fig:PL_s_mut3_expt}C and $\hat{Q}$ yields the black dashed line curve fit in Figure~\ref{fig:PL_s_mut3_expt}C. As expected, both parameter sets chosen along the curve that outlines the predicted relationship between $\beta_1$ and $\beta_2$ yield very close fits to the data.

\section{Discussion}\label{sec7} 

In the present work, we propose methods for assessing parameter identifiability and for learning identifiable parameter combinations based on \textblue{fluorescence microscopy measurements of protein dynamics and assuming that} a partial differential equations model \textblue{appropriately models these dynamics}. Here, we are specifically motivated by the recent discovery that RNA localizes together with RNA-binding proteins in L-body RNP granules during the development of frog oocytes \cite{neil2021bodies}. PTBP3 is a specific multivalent RNA-binding protein, for which protein dynamics are regulated by RNA-binding in L-bodies \cite{cabral2022multivalent}. Experimental measurements of PTBP3 dynamics are quantified using FRAP. \textblue{This is a commonly used technique to study protein dynamics in living cells, and is typically thought to investigate protein diffusion, as well as binding characteristics and connections between intracellular compartments \cite{ishikawa2012advanced}.} We model the recovery of protein fluorescence in these experiments using reaction-diffusion partial differential equations, characterized by the diffusion coefficient and the binding and unbinding rate parameters. The FRAP model we investigate here is a linear two-state PDE system, with a postbleach initial condition that we derive based on the square bleach spot used in the experiments in \cite{cabral2022multivalent}.  

We first sought out insights from application of established methods of parameter identifiability to our PDE model of protein dynamics during FRAP. In particular, we evaluated structural parameter identifiability, which is based on model structure alone, using the Fisher Information Matrix \cite{reid1977structural,cobelli1980parameter} and differential algebra approaches \cite{renardy2022structural}. Despite the simple linear reaction-diffusion structure of the model, we find that structural identifiability is either difficult or impossible to establish for the \textblue{time-series data extracted from} the PDE model using these methods. 

Practical parameter identifiability considers issues in parameter inference due to the noisy features of real data. We therefore use experimental datasets for wild-type and mutant PTBP3 protein dynamics from \cite{cabral2022multivalent} and our previously-developed deterministic parameter estimation pipeline in \cite{ciocanel2017analysis} to roughly inform parameter regimes of interest. Using synthetic FRAP data generated using these parameter regimes, we investigate methods of practical identifiability based on Bayesian inference and profile likelihoods for the FRAP model. We find that practical identifiability using Bayesian inference has a high computational cost, due to the MCMC sampling of the parameter space that is required. \textblue{In addition, both of} these methods suggest that certain parameters are practically unidentifiable, \textblue{however} it remains challenging to determine the parameter relationships that could be inferred based on the available FRAP data. Recent work on subdiffusive protein motion in FRAP has also shown that only some of the model parameters were able to be identified from FRAP data in certain regimes studied \cite{alexander2022inferences}.

Since the existing methods point to identifiability issues for the reaction rates in the FRAP PDE model, \textblue{we propose an alternative strategy for determining the }relationship between the kinetic rate parameters using synthetically-generated FRAP datasets and contour curves of the error function between data and simulated recovery curves for a range of binding and unbinding rate parameter choices. The framework we propose for identifying parameter combinations \textblue{builds on the calculation of profile likelihoods in \cite{eisenberg2014determining}} and involves constructing a transverse curve to the contour curves of the error function. We thus re-parameterize the PDE model of FRAP using the diffusion coefficient of the protein and a parameter that describes this transverse curve. Carrying out profile likelihoods for these parameters identifies the level curve on which the true parameters must lie. We demonstrate that this approach recovers the original \textblue{protein diffusion coefficient and the relationship between binding and unbinding rates} for synthetic datasets. \textblue{The method also} predicts the relationship between reaction rates for experimental FRAP data. \textblue{This prediction of the diffusion coefficient and of the relationship between binding and unbinding rates gives us insights into PTBP3 protein dynamics in this work. More broadly, this methodology can be used to understand how other protein components interact and bind with RNA in L-bodies. This has the potential to characterize the strength of binding affinities of the many protein components that assemble in the RNP granules in developing \textit{Xenopus laevis} oocytes and other biological systems.}

The pipeline we propose has the potential to extend to identifying parameter relationships in other PDE models of biological systems. However, the approach becomes more challenging for larger numbers of parameters that need to be identified. For the application motivating this work, we have used the simplifying assumption that the PTBP3 reaction uses a single binding site; this is appropriate for the mutant studied in Figure~\ref{fig:PL_s_mut3_expt}, which has a single RNA binding domain capable of binding to L-body RNA \cite{cabral2022multivalent}. For systems where multiple independent binding sites are appropriate, parameter identifiability and inference are likely more difficult to investigate due to the increased dimension of the parameter space. More generally, the specific insights we provide on parameter combinations that are identifiable in FRAP are dependent on the assumption that the reaction-diffusion model we use is appropriate. We have previously studied settings where active transport of proteins needs to be included and impacts parameter estimation \cite{ciocanel2017analysis}. Recent work has also shown that experimental FRAP data cannot distinguish between normal diffusive and subdiffusive motion in large regions of parameter space \cite{alexander2022inferences}. Future work could aim to develop broadly-applicable methods of structural and practical parameter identifiability for PDE models of fluorescence microscopy data.


\textbf{Acknowledgments}

Ding, Mastromatteo, and Reichheld were partially supported by the NSF under grant DMS-174429. Sandstede was partially supported by the NSF under grants DMS-2038039 and DMS-2106566. The experimental work was funded by R01GM071049 from the NIH to Mowry. 

\textbf{Data Availability}

The datasets generated during and/or analyzed during the current study are available from the corresponding author on reasonable request.

\section{Appendix: Derivation of the sensitivity equations}\label{sec:appendix_senseqs}

\textblue{We provide details on deriving the sensitivity equations associated with PDE model system~\eqref{eq:diff_pause}. Here, we show the derivation of the PDE equations for the sensitivities with respect to the diffusion coefficient $D$, since the other sensitivities can be derived in a similar way. }

\textblue{We first consider the PDE for the free protein concentration:}
\begin{align}\label{eq:diffonly}
\frac{\partial f}{\partial t} & = D \Delta f -\beta_2 f + \beta_1 c \,.
\end{align}
\textblue{We denote $f_D=\frac{\partial f}{\partial D}$ and $c_D=\frac{\partial c}{\partial D}$. We differentiate equation~\eqref{eq:diffonly} with respect to the diffusion coefficient $D$ and assume that the protein concentrations are smooth and thus have continuous partial derivatives. Applying the chain rule yields:} 
\begin{align}\label{eq:sens_diffonly}
\frac{\partial}{\partial D}\left(\frac{\partial f}{\partial t}\right)= 
\frac{\partial f_D}{\partial t} & = \frac{\partial}{\partial D}(D \Delta f) - \frac{\partial}{\partial D}(\beta_2 f) + \frac{\partial}{\partial D}(\beta_1 c) \nonumber \\
&= (f_{xx}+f_{yy}) + D \frac{\partial}{\partial D}\left(f_{xx}+f_{yy}\right) - \beta_2 f_D + \beta_1 c_D \nonumber \\ 
 &= \Delta f + D \Delta f_D - \beta_2 f_D + \beta_1 c_D \,.
\end{align}

\textblue{Similarly, consider the PDE for the bound protein concentration:}
\begin{align}
\frac{\partial c}{\partial t} & = \beta_2 f - \beta_1 c\,.
\end{align}
\textblue{Differentiating this equation with respect to $D$ yields the sensitivity equation:}
\begin{align}\label{eq:sens_D}
\frac{\partial}{\partial D}\left(\frac{\partial c}{\partial t}\right) = \frac{\partial c_D}{\partial t} & = \beta_2 f_D - \beta_1 c_D \,.
\end{align}

\bibliographystyle{unsrt}  

\bibliography{references} 

\end{document}